\documentclass[twocolumn,trackchanges]{aastex631}

\shorttitle{Circumbinary dust disk masses in Orion}

\graphicspath{{./}{figures/}}

\begin{document}

\title{The influence of tight binaries on proto-planetary disk masses}

\correspondingauthor{Kevin Flaherty}
\email{kmf4@williams.edu}

\author[0000-0003-2657-1314]{Kevin Flaherty}
\affil{Department of Physics and Astronomy, Williams College, Williamstown, MA, 01267, USA}

\author[0009-0009-5655-7696]{Peter Knowlton}
\affil{Department of Physics and Astronomy, University of Victoria, Victoria, Canada}

\author{Tasan Smith-Gandy}
\affil{Department of Astronomy and Department of Physics, Williams College, Williamstown, MA, 01267, USA}

\author[0000-0002-4803-6200]{A. Meredith Hughes}
\affil{Van Vleck Observatory, Astronomy Department, Wesleyan University, 96 Foss Hill Drive, Middletown, CT, USA}

\author[0000-0002-5365-1267]{Marina Kounkel}
\affil{Department of Physics and Astronomy, University of North Florida, 1 UNF Dr, Jacksonville, FL, 32224, USA}

\author[0000-0002-4625-7333]{Eric Jensen}
\affil{Department of Physics and Astronomy, Swarthmore College, 500 College Ave, Swarthmore, PA, 19081, USA}

\author[0000-0002-5943-1222]{James Muzerolle}
\affil{Space Telescope Science Institute, 3700 San Martin Dr, Baltimore, MD, 21218, USA}

\author[0000-0001-6914-7797]{Kevin Covey}
\affil{Department of Physics and Astronomy, Western Washington University, Bellingham, WA, 98225-9164, USA}

\begin{abstract}

Binary systems are a common site of planet formation, despite the destructive effects of the binary on the disk. While surveys of planet forming material have found diminished disk masses around medium separation ($\sim$10--100 au) binaries, less is known about tight ($<$10 au) binaries, where a significant circumbinary disk may escape the disruptive dynamical effects of the binary. We survey over 100 spectroscopic binaries in the Orion A region with ALMA, detecting significant continuum emission among 21 of them with disk masses ranging from 1--100 M$_{\oplus}$. We find evidence of systematically lower disk masses among the binary sample when compared to single star surveys, which may reflect a diminished planet forming potential around tight binaries. The infrared excess fraction among the binary sample is comparable to single stars, although the tight binaries without significant ALMA emission display tentative evidence of weaker 3-5$\mu$m excesses. The depletion of cold dust is difficult to explain by clearing alone, and the role of additional mechanisms needs to be explored. It may be the result of the formation pathway for these objects, systematic differences in intrinsic properties (e.g., opacity) or a bias in how the sample was constructed.

\end{abstract}


\section{Introduction} \label{sec:intro}


Most of the solar-type and lower mass stars reside in binary or multiple star systems \citep{raghavan2010,offner2023}. 
While planets are known to be highly abundant around single stars, less is known about planets in multiple systems. Circumbinary planets, aka Tatooine-like or P-type planets, in particular, are the least well characterized due to the challenges in detecting them, but progress has been made in recent years \citep{doyle2011,welsh2012}. Radial velocity and transit observations suggest that circumbinary planets are as frequent as planets around single stars \citep{armstrong2014,martin2019}. Both circumbinary and circumstellar planets tend to be coplanar with the binary orbit \citep{li2016,dupuy2022,kostov2020,martin2019,schwamb2013,orosz2019,orosz2012}, and circumbinary planets tend to be Saturn-sized or smaller \citep{orosz2012b,schwamb2013}, although samples are limited. 

This high frequency of circumbinary planets exists despite the destructive effects of a binary on a disk. Binaries are expected to clear out material between roughly 1/3rd and 3 times the semi-major axis of the binary \citep{artymowicz1994,miranda2015,hirsh2020}, with the exact boundaries depending on the mass ratio and eccentricity of the binary. This leads to the outer truncation of circumstellar disks, and the inner clearing of circumbinary disks. 

Truncation of the outer disk can change the disk evolution timescale and create conditions that are harmful for planet formation. The viscous timescale is proportional to the outer disk radius, and an outwardly truncated circumstellar disk will evolve on a faster viscous timescale, as well as through enhanced photoevaporation \citep{rosotti2018} and enhanced dust radial migration and depletion \citep{zagaria2021a}. This truncation is seen in ALMA and near-infrared imaging even out to binary separations of $>$few thousand au, where the largest disks are seen around single stars \citep{pascucci2008,cox2017,zurlo2021}. Disk fluxes as measured by ALMA and in the near-infrared are lower around binaries, consistent with the decrease in disk mass associated with truncation \citep{cox2017,zurlo2021,zagaria2021b,akeson2019}. This truncation of the radius and mass likely happens very quickly after the binary forms as suggested by the fact that the fraction of disk systems that are binaries is consistent between older ($>$5 Myr) and younger (1 -- 2 Myr) clusters \citep{barenfeld2019}. 

Clearing of the inner disk can also influence the evolution of the disk. Just beyond the inner edge of the circumbinary disk, eccentric motion induced by the binary \citep{munoz2020b,ragusa2020,lubow2022,marzari2012} can enhance planetesimal collisions \citep{marzari2008,meschiari2012,paardekooper2012} and turbulent motion in this region induced by the binary can limit pebble accretion \citep{pierens2021}, both of which would limit planetesimal growth. On the other hand, it may be that angular momentum injected into the circumbinary disk by the binary slows viscous evolution, helping to maintain a high surface density later in a disk's life \citep{vartanyan2016}, and thus extending the epoch of planet formation.





Most studies have focused on infrared emission, which traces circumstellar material close to the individual stars, finding depletions and accelerated evolutionary timescales for binaries with separations less than 40 au \citep{duchene2010,cieza2009,kraus2012}. The typical emitting region of infrared and optical disk diagnostics ($\sim0.1$ -- 1 au) would be cleared out by binaries with semi-major axes $<$10 au and the clearing of material close to the stars is consistent with the lack of circumstellar planets \citep{moe2019}. But the clearing of circumstellar material does not rule out circumbinary planets as a large reservoir of cold material at $>$30 au may still be present. Sub-mm observations can directly trace the bulk of the planet forming material, but studies of circumbinary disks are  limited due to the fact that the tight binaries that would host circumbinary disks are intrinsically rare \citep{raghavan2010,Kounkel2019,offner2023} and because expensive high resolution spectroscopic surveys are needed to identify  spectroscopic binaries. Despite these limitations progress has been made. High-resolution sub-mm observations of DF Tau\citep{grant2024,Kutra2025} find circumstellar disks around each component of the 14 au binary, but no evidence of a circumbinary disk. The Taurus survey of \citet{harris2012} only included 5 detections for systems with semi-major axes $<$10 au, which \citet{akeson2019} expand upon with deeper ALMA observations, detecting disks around an additional 6 spectroscopic binaries. These studies find disk masses around tight binaries that are similar to those of single stars, but with limited statistics. 


The limited knowledge of disk properties around tight binaries makes it difficult to assess how strongly the central binary influences the disk and the planet formation process. Here we take advantage of large-scale spectroscopic surveys to expand the sample of systems observed for disk masses by an order of magnitude. In section~\ref{sec:data} we describe the 130 spectroscopic binaries identified in Orion, and our ALMA observations tracing their disk mass. In section 3 we discuss the finding of a depleted disk mass among these systems and in section 4 we discuss the possible explanations for this result.

\section{Data \& Methodology} \label{sec:data}
\subsection{Circumbinaries in Orion}
\subsubsection{Base Spectroscopic Sample}
We focus on binaries with small separations, as identified through spectroscopic surveys. We draw spectroscopic binaries from two surveys of Orion, \citet{Kounkel2016} and \citet{Fernandez2017}. \citet{Kounkel2016}, using and building on multi-epoch MMT-Hectochelle ($\lambda\sim5150$ -- $5300$\AA, R $\sim35,000$) and Magellan-MIKE ($\lambda\sim5150$ -- $5210$\AA, R $\sim18,000$) observations of 727 objects from \citet{tobin2009}, identify binaries based on radial velocity (RV) variability. \citet{Fernandez2017}, using observations of 2700 pre-main sequence stars from the Sloan Digital Sky Survey APOGEE ($\lambda\sim1.5$--$1.6\mu$m, R $\sim$ 22,500) INfrared Spectroscopy of Young Nebulous Clusters program (IN-SYNC) survey of the Orion A molecular cloud \citep{dario2016}, look for double-lined spectroscopic binaries. Our sample includes 65 sources from \citet{Kounkel2016} and 65 sources from \citet{Fernandez2017}, for a total of 130 spectroscopic binaries. 


\begin{figure}
    \centering
    \includegraphics[width=1\linewidth]{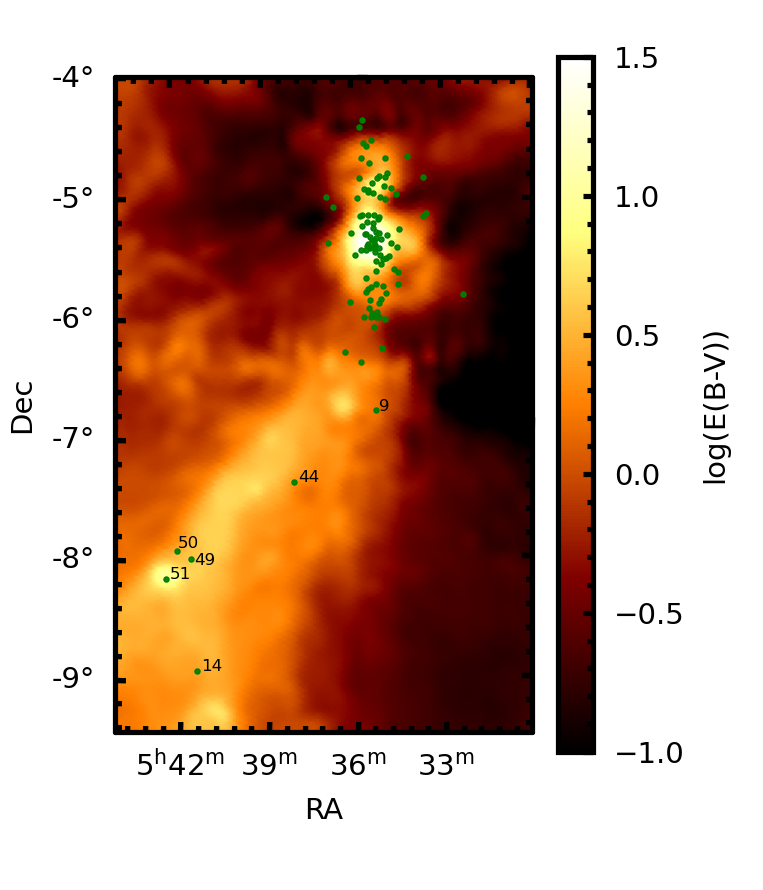}
    \caption{Location of our tight binary targets on a map of dust extinction \citep{schlegel1998}. Green circles indicate the targets of our ALMA observations that made it through the cluster membership cuts (section~\ref{sec:cleaning}). The majority of our targets are located near the Orion Nebula Cluster, with some spread along the Orion A cloud; the more isolated targets in Orion A are marked with their ID numbers.}
    \label{fig:target_map}
\end{figure}

\subsubsection{ALMA Survey}
ALMA observations of this sample (2017.1.01639.S, PI: K. Flaherty) were collected on 16 Aug 2018 and 23 Aug 2018, with $\sim$40 second on-source integrations. Baselines ranged from 15 m to 500 m. Bandpass and flux calibration was done with observations of J0522-3627, while phase calibration was done with J0542-0913. Three 2 GHz wide continuum spectral windows, centered at 249.46 GHz, 251.96 GHz, and 264.96 GHZ were utilized. A fourth spectral window was centered at 267.52 GHz, with a bandwidth of 937.5 MHz and a channel spacing of 1128.906 kHz, corresponding to 1.3 km s$^{-1}$, in order to cover the HCO$^+$(3--2) transition.


The raw data were pipeline calibrated by NRAO staff using CASA (version 5.4.0). Each field was cleaned using \texttt{tclean} in CASA \citep{casa}, using natural weighting, and 5000 iterations, resulting in a typical beam size of 1$\farcs$03$\times$0$\farcs$73 and a typical rms of 0.15 mJy/bm. Four fields have rms values much larger than this, and are likely contaminated by cloud emission. Continuum detections were defined as having peak SNR$>$5$\sigma$ and by this definition we detect continuum emission in 31 sources. We detect HCO$^{+}$ emission in five cluster members, as discussed in more detail in section~\ref{sec:hco_gas_mass}. The uncertainties on the continuum emission do not include the $\sim$10\% uncertainty on the absolute flux calibration. When no continuum emission is detected, we combine all of the spectral windows to place an upper limit on the continuum emission. 

\begin{figure*}
    \includegraphics[scale=2]{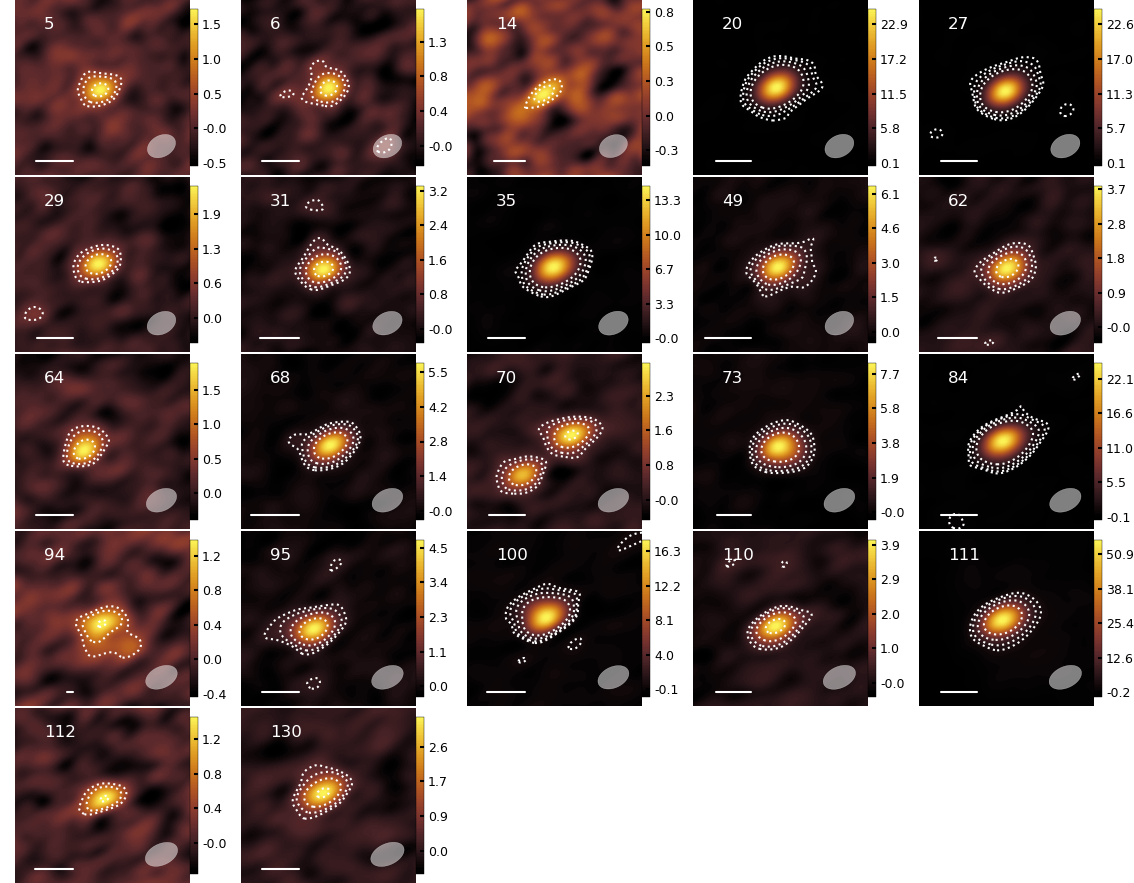}
    \caption{Continuum images, centered on the detections, of cluster members. Each panel is 6\arcsec on a side, with the scale bar in units of mJy bm$^{-1}$.} Dotted contours are at 3, 5, 10, and 20$\sigma$. The horizontal line marks 500 au, while the ellipse in the lower right is the beam size, and the numbers in the upper left are cluster ID numbers.
    \label{fig:cont_images}
\end{figure*}

Figure~\ref{fig:cont_images} shows the detected continuum emission from cluster members (see discussion below about interlopers). All of the images show single unresolved point sources with the exception of sources 70 and 94. In source 70 two detections are found within 2$\arcsec$ of each other, while for source 94 there is a faint extension in the south-west direction. When a source is detected, we fit a point source model to the visibilities in the continuum spectral windows using \texttt{uvmodelfit} to derive the flux. It is not surprising that the sources are unresolved given that the $\sim$1$\arcsec$ resolution corresponds to $\sim400$ au, and that dust disks are typically much smaller than 400 au \citep[e.g.,][]{eisner2018}.  

Appendix~\ref{sec:full_data} contains more complete data on the full sample. Infrared photometry is taken from 2MASS \citep{Skrutskie2006}, WISE \citep{Wright2010}, and Spitzer Space Telescope \citep{Megeath2012} surveys. {\it Gaia} DR3 observations \citep{GAIA2021} are used for the parallax, which is available for all but four objects. Effective temperature ($T_{\rm eff}$), $\log g$, and YSO class are taken from \citet{Kounkel2016}, \citet{Kounkel2019}, and \citet{sizemore2024}. Effective temperatures, which are biased towards the primary star, are available for all but eight of our targets, and $\log g$ values are available for all but 28 of our targets. Derivation of these quantities is described in \citet{Cottaar2014}, \citet{Kounkel2019}, and \citet{sizemore2024}. Other properties of the binary are more limited --- e.g., only 12 objects have measurements of the mass ratio \citep{Kounkel2019} --- and are not used in our analysis. Further work is needed to characterize the underlying binaries in our sample to understand if properties such as the exact binary separation or the mass ratio are directly related to the presence of cold dust. 


\subsection{Removing Interlopers\label{sec:cleaning}}
The original sample was drawn from different surveys that used different criteria for assessing cluster members. Fibers for the IN-SYNC survey were assigned with preference for objects that were previously known members of optical surveys of the ONC, or had X-ray emission or an infrared excess \citep{dario2016}. Fibers for the Magellan and MMT surveys were assigned to lightly extinguished sources based on their position in a JHK color-magnitude diagram, as well as sources that exhibited an infrared excess in {\it Spitzer Space Telescope} observations \citep{tobin2009}. These selection criteria are not fool-proof, and foreground or background objects can leak in to the sample.  

With {\it Gaia} DR3 \citep{GAIA2016,GAIA2021} we have more detailed information on the parallaxes and proper motions of our targets than was available when the sample was originally constructed, allowing for a more detailed evaluation of cluster membership. Follow-up spectroscopy by \citet{Kounkel2019} has added RV measurements, as well as T$_{\rm eff}$ and $\log~g$ values for much of the sample. Also, \citet{Kounkel2018} have performed a detailed analysis of the kinematics of the Orion star-forming region, providing better information on the range of parallaxes, proper motions, and radial velocities associated with different clusters.

Based on this information, we apply the following cuts:
\begin{itemize}
    \item {\it Off-center cut:} We exclude any source that was detected at a position that was off-center from the phase center of an observation by greater than 1$\arcsec$. This excludes 11 sources.
    \item {\it Parallax cut:} We exclude any source with parallax $p<$ 1.5 mas or $p>$ 3.5 mas. Of the 104 sources with good parallax measurements (RUWE$<1.4$), 9 sources are excluded based on this criterion. 
    \item {\it Proper Motion cut:} We exclude any source that fails a proper motion cut ($|\mu_{\alpha}|>$5 mas yr$^{-1}$ or $|\mu_{\delta}|>$5 mas yr$^{-1}$). This removes an additional 11 sources.
\end{itemize}
Our final clean sample consists of 107 sources, 21 of which have significant ALMA emission (Table~\ref{table:continuum_data}). 

{\it Off-center cut:} The first cut removes interlopers that fall in the field of view of our observations, but are not the intended target. These could be single stars with disks within the cluster, or background galaxies. Detections within 1$\arcsec$ of the phase-center are kept since they fall within the fiber diameter of the spectroscopic observations; Hectochelle has 1$\farcs$5 diameter fibers, the MIKE fiber system on Magellan has 1$\farcs$2 diameter fibers, and APOGEE has 2$\arcsec$ diameter fibers. If an off-center source is detected within a field, but no source is detected within 1$\arcsec$ of the phase center, we include the phase center as a non-detection. In terms of constructing a sample of tight binaries, we want to include the upper limit associated with the tight binary, while excluding the interloper. The off-center detections are included in Table~\ref{table:continuum_data_excluded}, and shown in Figure~\ref{fig:cont_images_excluded}, along with any $T_{\rm eff}$, $\log g$, and parallax data available from \citet{Kounkel2019} and \citet{GAIA2021}, for completeness.

{\it Parallax cut:} The second cut uses {\it Gaia} observations to filter out foreground and background sources. We restrict this cut to sources with Renormalized Unit Weight Error (RUWE) less than 1.4. The RUWE characterizes how much the image center deviates from the astrometric model and this boundary restricts our cut to sources with accurate astrometry \citep{Lindegren2021}. \citet{Kounkel2018} find that the Orion complex, and particularly the ONC region where our sample is concentrated, is made up of multiple groupings with parallaxes ranging from 2.0 to 3.5 mas. The ONC is located at p$\sim$2.5 mas, while the Orion D region, which is along the line of sight to the ONC, has p$\sim$2.8 mas. Our boundaries are defined to broadly encompass this range of regions. Of the 104 sources with RUWE$<$1.4, nine sources fail this cut. We do find that many of the sources with 1.4$<$RUWE$<$2 have parallaxes consistent with the cluster, while sources with larger RUWE have discrepant parallaxes. As a result, for sources without a parallax measurement, or with RUWE$>$2, we assume $p$=2.55 mas, equal to the average for the ONC \citep{Kounkel2018}.

{\it Proper Motion Cut:} The third cut uses proper motion measurements from {\it Gaia}, for those sources with RUWE$<$1.4, with boundaries again taken from \citet{Kounkel2019}. This removes an additional eleven sources. Note that we do not remove sources with RUWE$>$1.4, as the deviation from the astrometric model that causes the large RUWE may be due to the binary.

\begin{figure*}
\center
    \includegraphics[scale=1.2]{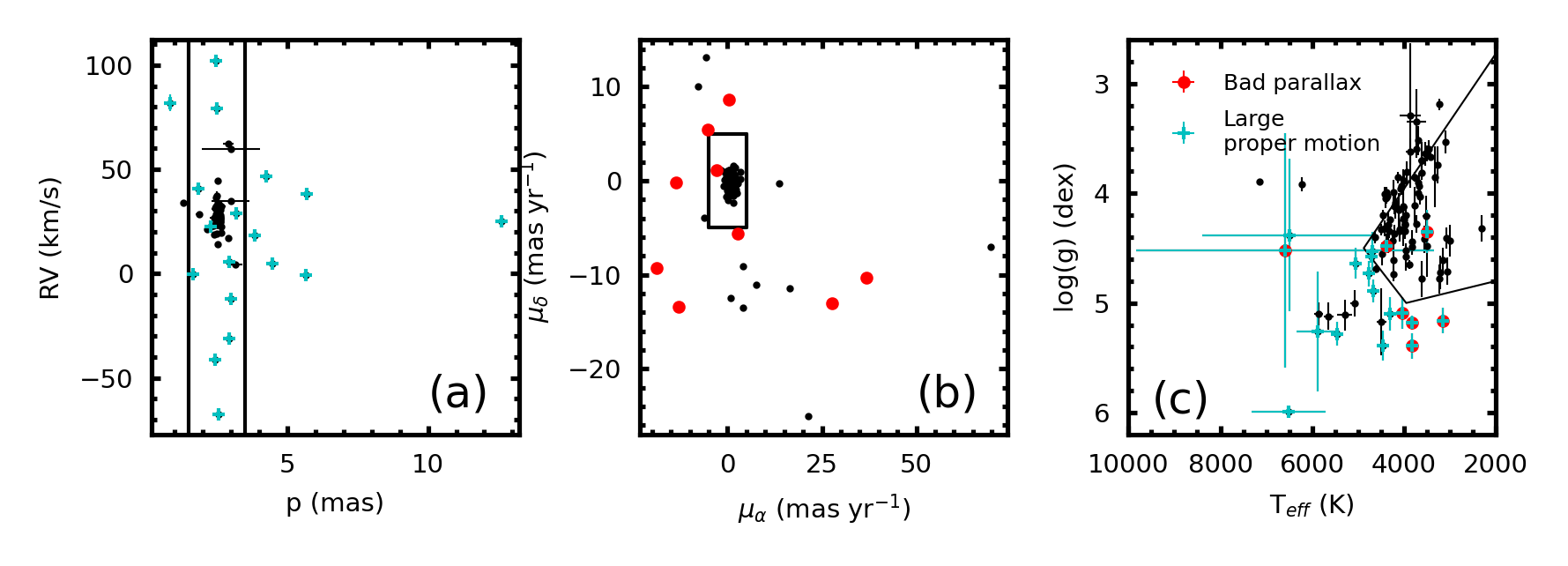}
    \caption{Membership diagnostics among the sources in our survey. We employ cuts on parallax ($p$) and proper motion ($\mu_{\alpha}$, $\mu_{\delta}$) to remove interlopers; these boundaries are indicated by solid black lines. RV is not used to remove non-cluster members as it may be biased by the presence of a binary. Many of the sources identified as interlopers by their discrepant parallax and/or proper motion values also fall outside the pre-main sequence region in the $\log~g$ vs $T_{\rm eff}$ diagram, enclosed by the solid lines, confirming their status as interlopers. \label{figure:source_cuts}}
\end{figure*}

{\it Overlap in source cuts:} Figure~\ref{figure:source_cuts} shows the distribution of parallaxes, RV\footnote{Throughout this paper RVs are in a Barycentric reference frame.}, and proper motions for our sample, delineating sources retained in our sample and sources that were excluded based on the second and third criteria above. There is substantial overlap among the second and third criteria; half of the sources with discrepant parallaxes also have discrepant proper motion values. Figure~\ref{figure:source_cuts} also shows T$_{\rm eff}$ vs $\log~g$ for our sources. The sources cut by our criteria fall outside the pre-main sequence region defined by \citet{Kounkel2019} in T$_{\rm eff}$ vs $\log~g$, confirming our choice of rejection conditions. Also, none of the excluded sources are detected in ALMA, consistent with the fact that they are foreground or background objects that are older that the Orion cluster members, and therefore less likely to have cold dust emission.

Figure~\ref{figure:source_cuts} also demonstrates why we do not use radial velocity to find cluster members. There is a large spread in RV, likely because many of our sources are defined by their large RV variability, which will bias the average RV. Many of the sources with discrepant RV values have parallaxes and proper motions that fall within the acceptable range, while the sources with the most discrepant RV values fail another cut. 

Among the 11 off-center detections, which are not included in Figure~\ref{figure:source_cuts}, parallaxes are available for seven sources, with six of them (17B, 70B, 107C, 113B, 116B, 130B) having parallaxes consistent with the membership in the cluster. Among these, three (70B, 116B, 130B) have proper motion measurements that are also consistent with membership in the cluster, and the proper motion for all of these off-center detections are significantly different from that of the target at the phase center. This indicates that they are cluster members but are unlikely to be distant tertiary companions to the target spectroscopic binary, although further observations are needed to fully rule this out.

\subsection{Dust mass}
Dust masses are derived assuming optically thin dust emission. 
\begin{equation}
    M_d = \frac{F_{\nu}d^2}{\kappa_{\nu} B_{\nu}(T_d)}\label{eqn:mass}
\end{equation}
where $M_d$ is the disk dust mass, $F_{\nu}$ is the measured flux, $d$ is the distance, and $\kappa_{\nu}$ is the dust opacity. We assume that the dust temperature $T_d$ is 20 K and $\kappa_{\nu}=2.5$ cm$^{2}$ g$^{-1}$, in line with previous studies \citep{andrews2020,miotello2023}.  Disk mass uncertainties include the flux uncertainty and, where available, parallax uncertainty, but do not include the $\sim$10\% uncertainty in the absolute flux calibration \citep{Butler2012}, or any systematic uncertainty associated with our choice of $T_d$ and $\kappa_{\nu}$. Among the detected systems we measure a median dust mass of 11.6 M$_{\oplus}$, with a maximum of 95.5 M$_{\oplus}$, with typical 5$\sigma$ upper limits among the non-detections of 1-2 M$_{\oplus}$. 

Caution should be used in interpreting the dust masses as the total reservoir of material in the disk as equation 1 may underestimate the total available mass within the disk \citep{ballering2019,ribas2020,rilinger2023}. When modeling the full spectral-energy distribution, \citet{ballering2019} and \citet{rilinger2023} find that many disks, including sources in Orion and other star-forming regions, are optically thick and have larger disk masses than estimated from equation~\ref{eqn:mass}. Despite these complications, we have chosen common assumptions to allow for a direct comparison with the results of other studies, since our goal is to provide a comparison between circumbinary disks and disks around single stars. More detailed analysis is needed to fully characterize the amount of planet-forming material around these systems.

\subsection{HCO$^+$ gas mass\label{sec:hco_gas_mass}}
In addition to the continuum observations, we devoted one spectral window to cover HCO$^+$, in order to search for gas emission. We avoid CO emission as it is likely to be contaminated by emission from the surrounding molecular cloud. We search for HCO$^+$ among those sources with continuum emission, since only the most massive disks are likely to have HCO$^+$ emission. Tables \ref{table:hco} and \ref{table:hco-excluded} list the results for the clean and excluded samples respectively. For detections integrated fluxes are derived from Gaussian fits to the moment 0 maps, excluding channels with obvious cloud contamination, while for non-detections upper-limits are derived assuming unresolved emission spread over 7 channels.

Assuming optically thin emission and local thermodynamic equilibrium, the gas mass can be related to the integrated HCO$^+$ flux via:

\begin{eqnarray}
    M = \frac{4\pi F m d^2}{h \nu_0 A_{ul} X_u X_{HCO+}}\\
    X_u = \frac{2J_u+1}{Q(T)}\exp^{(-E_u/kT_{ex})}
\end{eqnarray}
where $m$ is the molecular mass, $F$ is the integrated flux, $A_{ul}$ is the Einstein A value for the J=3-2 transition, $\nu_0$ is the rest frequency of this transition, $X_u$ is the fraction of molecules in the upper rotational state, $E_u$ is the energy of this upper state, and $T_{ex}$ is the excitation temperature. The parameter $X_{HCO+}$ is the abundance of HCO$^+$ relative to the total gas mass, which we take to be 10$^{-9}$. For simplicity we assume all targets have the same HCO$^+$ abundance, although we note that $X_{HCO+}$ can depend on the external radiation field \citep{gross2025} and the CO abundance \citep{walsh2012,seifert2021}, both of which may vary across our sample \citep[e.g.,][]{zhang2021,anania2025}. In the case of a linear rotator like HCO$^+$, with $kT_{ex}>hcB_0$, the partition function $Q(T)$ can be approximated as $Q(T) = kT_{ex}/hcB_0$, where $B_0$ is the rotation constant \citep{goldsmith1999}. Molecular data ($A_{ul}=1.4757\times10^{-3}$ s$^{-1}$, $\nu_0=267.558$ GHz, $B_0=44594.4\times10^{6}$ MHz, $E_u=25.68$ K) are taken from the Leiden LAMBDA database \citep{schoier2005}.

We detect HCO$^+$ emission from six sources, with gas masses ranging from 2.0$\pm$0.4 M$_{\rm jup}$ to 15.1$\pm$0.8 M$_{\rm jup}$. The HCO$^+$ detections are found among the disks with the largest dust masses, as expected given that HCO$^+$ has a low abundance and would be easiest to detect in the most massive disks. Gas to dust mass ratios range from 10 -- 300 among this sample. 

HCO$^+$ spectra are shown in Figure~\ref{fig:hcoplus}, along with parametric model predictions, the details of which are described in Appendix~\ref{sec:hcoplus_models}. The models are not quantitatively fitted to the data, but are illustrative of what would be expected for standard disk assumptions (e.g., power law radial temperature and surface density profiles, moderate inclination, Keplerian motion). 
Sources 20 and 27 exhibit broader spectra than the fiducial models, which might be due to a combination of high inclination and higher stellar mass than assumed. It is also possible that the gas is concentrated in a ring at the inner edge of the disk, similar to the dust emission around CS Cha \citep{kurtovic2022}, V4046 Sgr \citep{martinez-brunner2022}, and V892 Tau \citep{long2021}, which would amplify the high velocity emission. We note that, as discussed in more detail below, the central clearing may be larger than predicted from gravitational clearing by the binary. 
The systematic velocities of the disks are consistent between the ALMA observations and the previous optical spectroscopy, with the notable exception of source 27, where the ALMA data is blueshifted by $\sim$7 km s$^{-1}$ relative to the optical spectra. Further work is needed to fully understand these discrepancies and the underlying HCO$^{+}$ structure in these systems.

\begin{figure*}
\centering
    \includegraphics[scale=1.5]{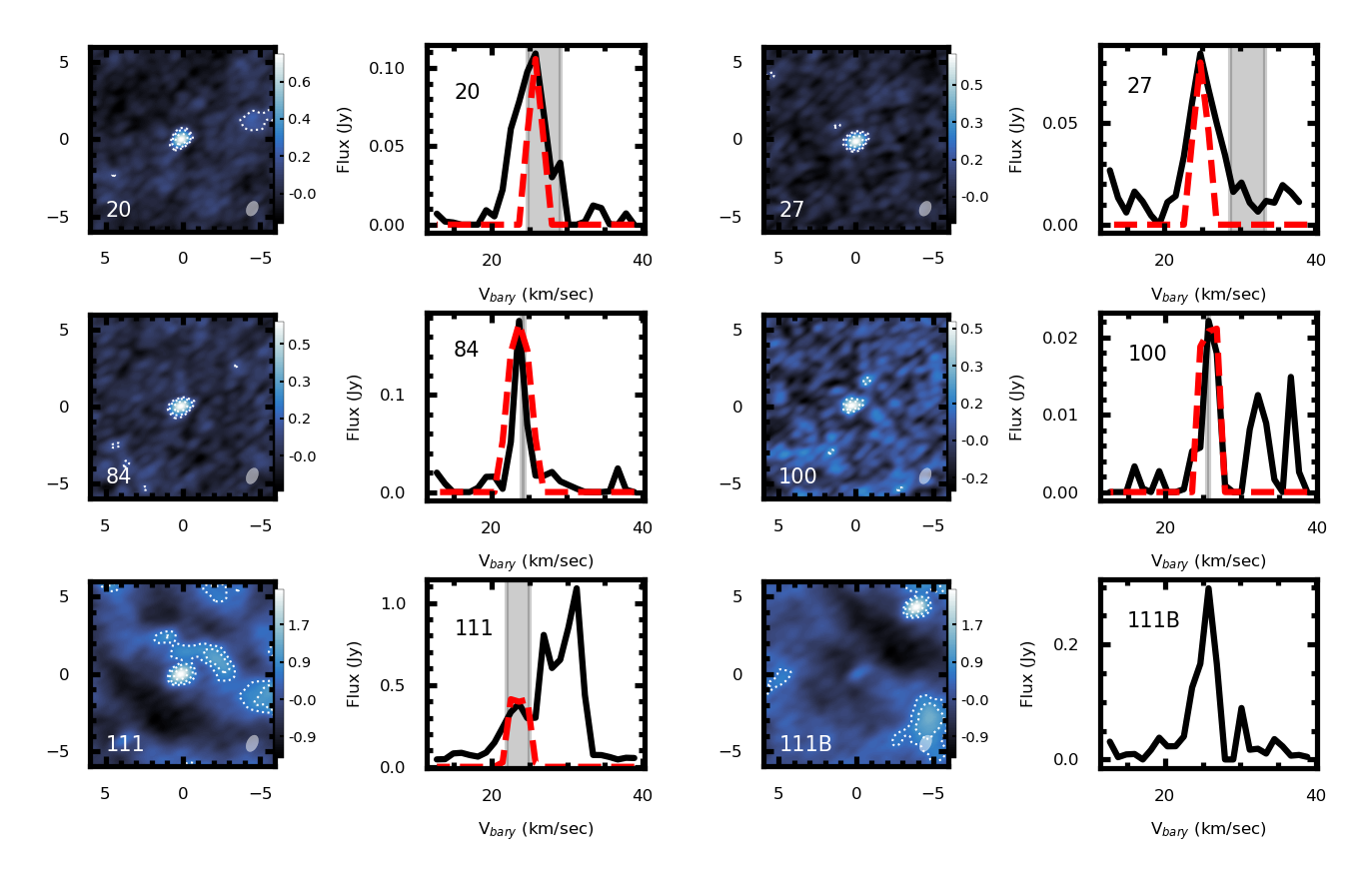}
    \caption{HCO$^+$ detections, including moment 0 maps (in units of Jy bm$^{-1}$ km s$^{-1}$), and spectra (in the barycentric reference frame, and in units of Jy)). Cloud contamination is present in some of the spectra at v$_{\rm bary}\sim$30 km s$^{-1}$. Dotted contours in the moment 0 maps are at 3, 5, 10, and 20$\sigma$. Grey bands indicate the $\pm$1 $\sigma$ ranges for the stellar velocities as measured in optical spectra by \citet{Kounkel2019}. Red dashed lines show model disk emission for comparison, generated using parametric models as described in Appendix~\ref{sec:hcoplus_models}. }
    \label{fig:hcoplus}
\end{figure*}



\section{Results}

\subsection{Evidence for circumbinary disk depletion}
\begin{figure*}
\center
\includegraphics[scale=1.]{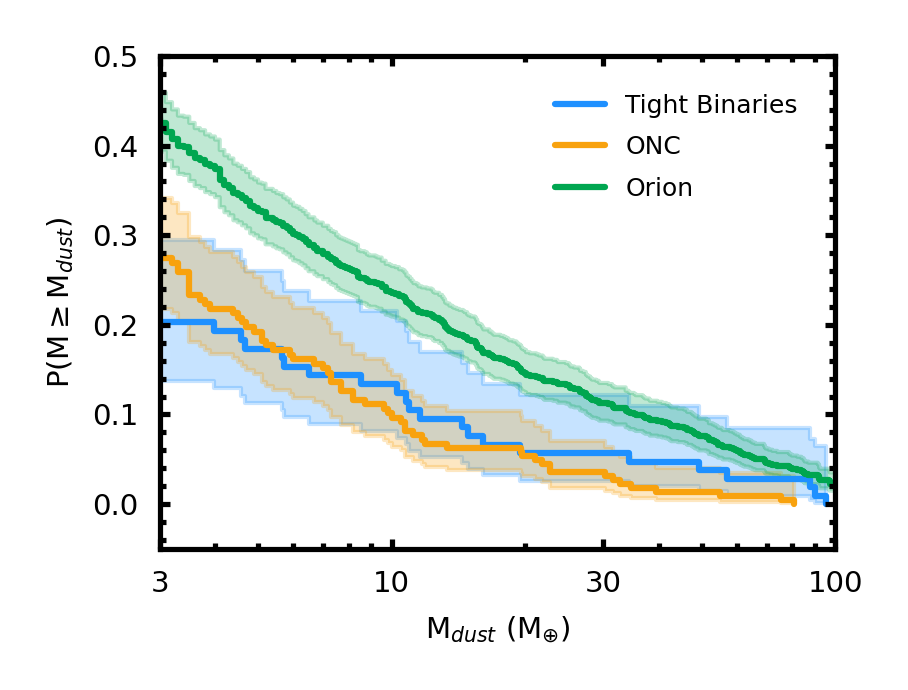}
\includegraphics[scale=1.]{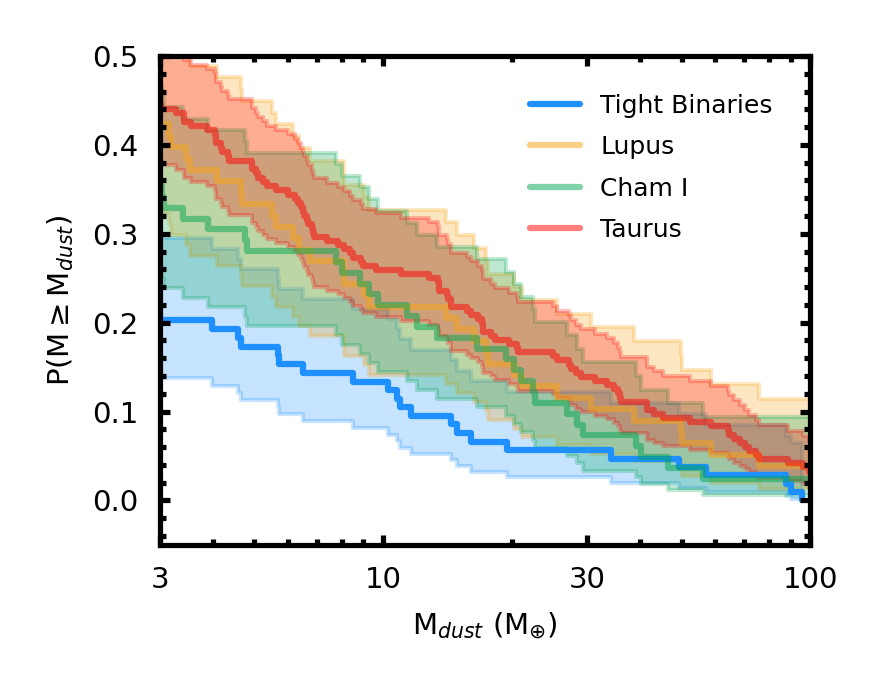}
\caption{Disk dust mass distribution for the tight binaries from our sample  and as compared with the ONC \citep{eisner2018} and Orion \citep{vanTerwisga2022} (left panel) and the young clusters Lupus, Cham I , and Taurus (data compiled by \citet{manara2022}, right panel). The disk masses around tight binaries are significantly depleted relative to the single stars in Orion and Lupus/Cham I/Taurus, and are comparable to those around stars in the ONC, which exhibits a depletion from photoevaporation from nearby massive stars. \label{figure:mass_dist}}
\end{figure*}

Among the 107 cluster members in our sample, we detected significant ($>$5$\sigma$) dust disk emission from 21. Figure~\ref{figure:mass_dist} shows the dust mass distribution as inferred using the Kaplan-Meier estimator, implemented in the \texttt{lifelines} package \citep{davidson_pilon_cameron2022}. For comparison, we also show the dust mass distributions for the ONC \citep{eisner2018} and the larger Orion cloud \citep{vanTerwisga2022}, as well as the young clusters Lupus \citep{ansdell2016,ansdell2018}, Cham I \citep{pascucci2016,long2018}, and Taurus (\citet{andrews2013,hardy2015,akeson2014,akeson2019,ward-duong2018,long2019}, as compiled by \citet{manara2022}). 

The tight binary disk mass distribution is consistently lower than the Orion sample and the Lupus/Cham/Taurus clusters, by a factor of $\sim2$, similar to the depletion seen among binaries with separations of 10 -- 100 au \citep{harris2012}. To compare the samples in a more statistically robust manner we use the log-rank correlation test. To do so we follow \citet{feigelson1985} and \citet{vanTerwisga2022} in subtracting our data from a constant, since \texttt{lifelines} implements this test for right-censored data, while our upper limits are left-censored data. We find significant differences between the tight binaries sample and that of Lupus, Chameleon, Taurus, and the Orion samples (all p$<$0.005). The lack of a significant difference between the tight binary sample and the ONC sample is discussed in detail below (Section~\ref{sec:photoevaporation}).






\subsection{Inner Disk Emission}
\begin{figure*}
\center
\includegraphics[scale=1.]{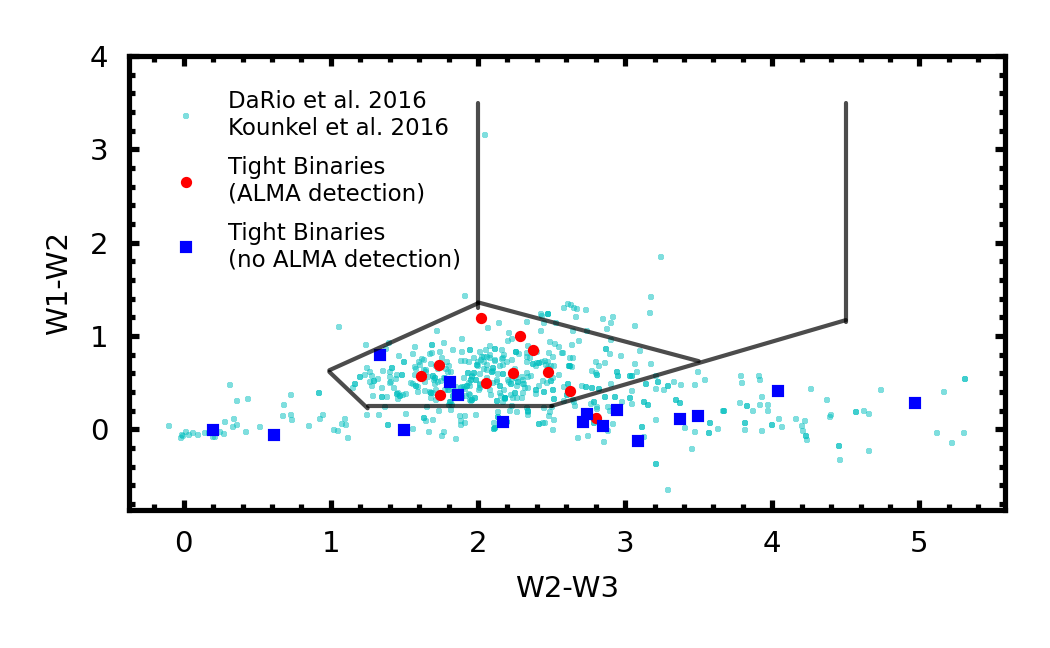}
\includegraphics[scale=1.]{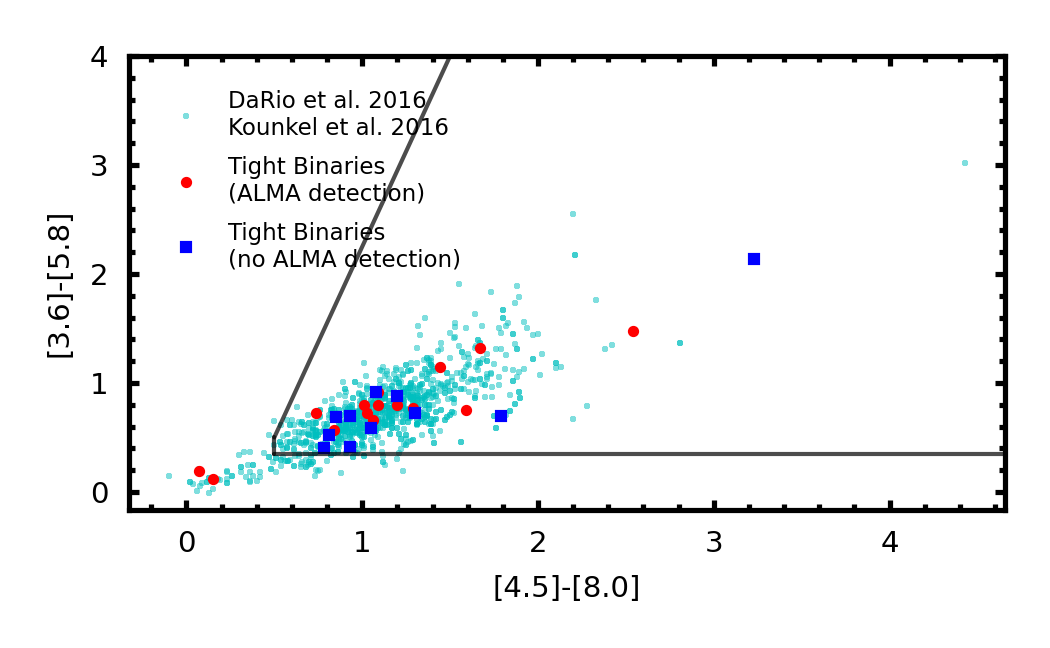}
\caption{Infrared color-color diagrams using photometry from WISE (left) or Spitzer (right). In the left panel the polygon indicates the region containing class II sources, while the extensions upwards demarcate the region with class I sources \citep{Fischer2016}. In the right panel the black lines delineate sources with a significant infrared excess \citep{Gutermuth2008}. Inner disk excess emission is most evident in the longer-wavelength observations of WISE and Spitzer, in particular beyond $\sim$5$\mu$m (W2-W3, [3.6]-[5.8], [4.5]-[8.0]). \label{figure:ir_ccds}}
\end{figure*}

We now turn to the inner disk, as traced by infrared emission, to see if it reflects the depletion seen in the outer disk. Figure~\ref{figure:ir_ccds} shows infrared color-color diagrams for the tight binary sample as compared to the parent samples from \citet{dario2016} and \citet{Kounkel2016}, using disk boundaries as defined by \citet{Fischer2016} and \citet{Gutermuth2008} for WISE and Spitzer, respectively. 

We find that the frequency of infrared excesses is similar between the tight binary sample and the parent samples (Table~\ref{table:ir_excess}), indicating that the binary does not completely clear the inner disk. Splitting apart binary systems with or without ALMA detections, we find that nearly all of the sources with ALMA detections have an infrared excess (Figure~\ref{figure:ir_ccds}). Binary systems without ALMA detections show long-wavelength excesses (e.g., [4.5]-[8.0], [3.6]-[5.8], W2-W3) comparable to systems with ALMA detections, but have systematically lower W1-W2 excesses. These objects have colors similar to those of transition disks,\footnote{Some objects have colors consistent with star-forming galaxies \citep[W2-W3$>$3, W1-W2$<1$;][]{Fischer2016}, but all of these objects have parallaxes measured by {\it Gaia} that are consistent with the Orion cluster, ruling them out as extra-galactic contaminants.} which may reflect the presence of some cool dust among all of these systems, but a lack of warmer dust that emits at 3-5$\mu$m among the binary systems without ALMA detections. As discussed below this depletion of warmer dust may reflect the lack of replenishment of the inner disk in systems without large cold dust reservoirs. Detailed infrared spectral energy distribution (SED) modeling, and more detailed sampling of the infrared SED, is needed to verify and fully interpret the infrared colors among this sample.

\begin{deluxetable}{cc}
 \tablecaption{IR excess fraction\label{table:ir_excess}}
\tablehead{\colhead{Survey} & \colhead{Excess Fraction}}
\startdata   
\cutinhead{WISE}
This study & 48 $\pm$ 13\% (13/27)\\
\citet{dario2016} & 62 $\pm$ 4\% (299/486)\\
\citet{Kounkel2016} & 55 $\pm$ 7\% (66/120)\\
\cutinhead{Spitzer}
This study & 92 $\pm$ 19\% (24/26) \\
\citet{dario2016} & 84 $\pm$ 3\% (798/954) \\
\citet{Kounkel2016} & 88 $\pm$ 5\% (345/394) \\
\enddata
\end{deluxetable}

\section{Discussion}
In our survey of ALMA emission from spectroscopic binaries in Orion we find evidence of a systematic depletion in the cold dust emission compared to that around single stars. Many of the binaries still exhibit an infrared excess, with the binaries without ALMA detections showing evidence of weaker 3-5$\mu$m excesses. Here we explore possible explanations for these observations. 

\subsection{Causes of Circumbinary Disk Depletion}
\subsubsection{Gravitational Clearing of the Disk}
\begin{figure*}
\center
\includegraphics[scale=1.]{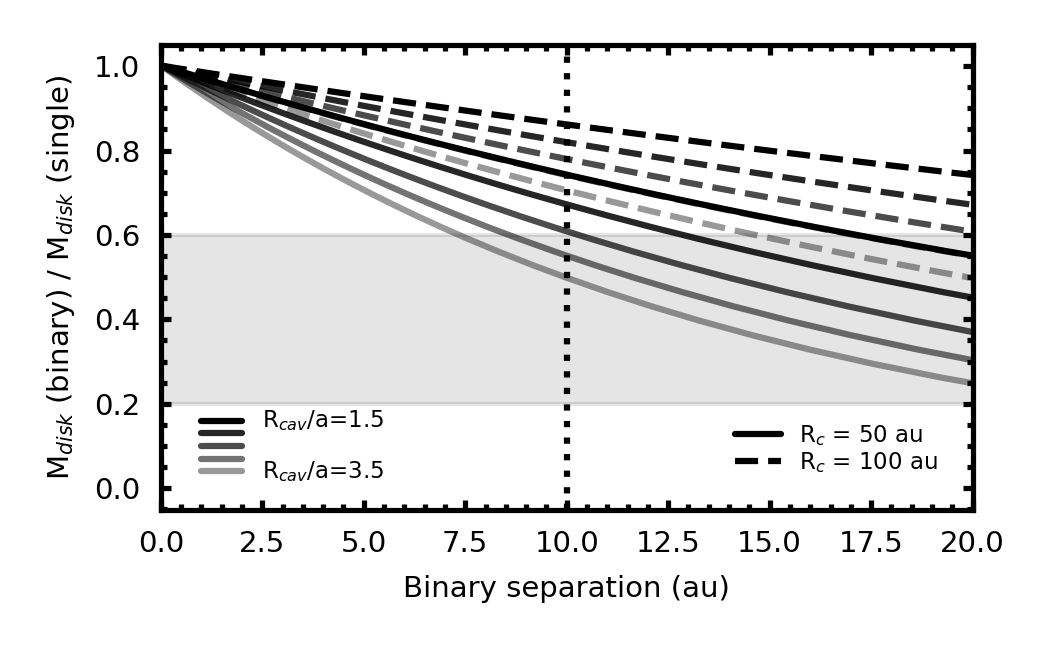}
\includegraphics[scale=1.0]{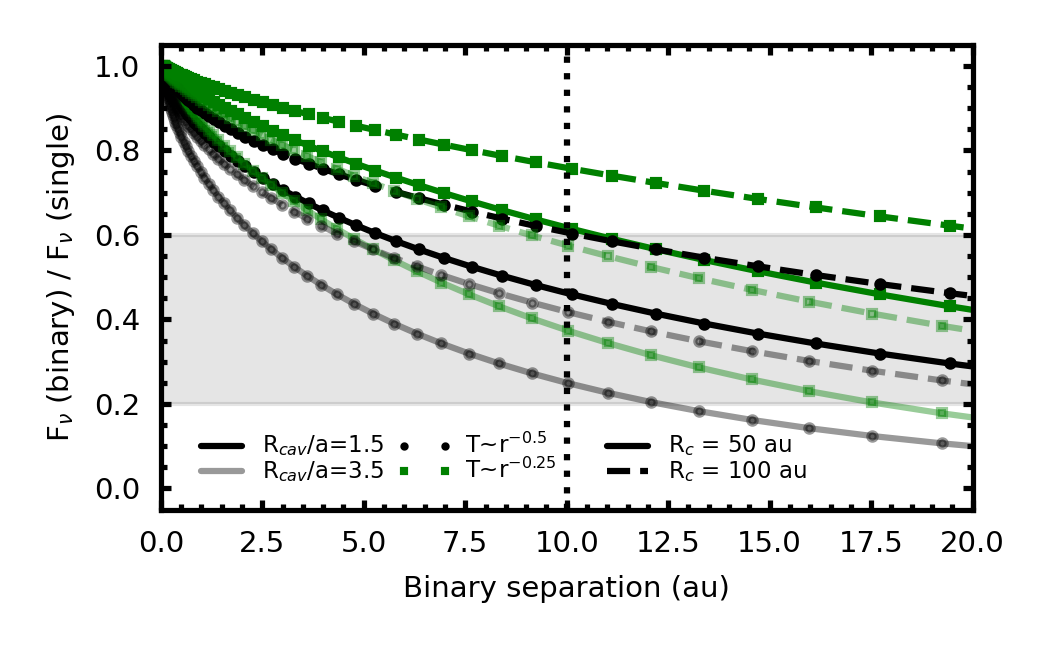}
\caption{A central binary will remove some material reducing the total disk mass (left panel) depending on the size of the disk (solid vs dashed lines) and the size of the cavity (gray-scale of the lines). The solid gray band represents our observed dust mass depletion. While the binary separations in our sample are not known, the spectroscopic resolution in the surveys used to define our sample limit the sample to separations $<$10 au, indicated by the vertical dotted line. By carving out the inner disk, the binary also changes the average temperature of the disk, which, in combination with the removed disk mass, reduces the observed dust disk flux (right panel). A central clearing associated with the binary can explain the observed disk only for relatively large binary separations in compact disks with large cavities and steep temperature profiles. \label{figure:mdisk_fraction}}
\end{figure*}

The primary effect of a binary on the disk is a clearing out of gas and dust from around the binary. Depending on the extent of the clearing, and the separation of the binary components, this clearing could explain the observed mass depletion. To quantify this effect, we consider the truncation of an otherwise smooth disk. We start with a self-similar functional form to the disk surface density ($\Sigma\sim(r/R_c)^{-\gamma}\exp((r/R_c)^{2-\gamma}$, \citealt{lynden-bell1974}), and remove  all of the dust within a radius corresponding to the size of the clearing ($R_{\rm cav}$). We consider different values of the extent of the clearing relative to the binary separation ($R_{\rm cav}/a=$1.5--3.5), as well as different values of the critical radius. We do not account for any dust pile-up or trapping at the outer edge of the binary-cleared cavity, which can also affect the continuum flux \citep[e.g.][]{Dullemond2018,pinilla2018}.  
Figure~\ref{figure:mdisk_fraction} shows the disk depletion for binaries with separations up to 20 au, although the velocity resolution of the spectroscopic observations used to define our sample limit our sample to separations $<$10 au \citep{Kounkel2019}. As expected, the size of the depletion increases as the binary separation, and the size of the cavity relative to the binary separation, increases. Our ALMA observations are consistent with circumbinary disk masses 20-60\% as large as around single stars. For binaries with separations $<$10 au, the depletion reaches 50\% for compact disks ($R_c=50$ au) around wide binaries ($a\sim10$ au) with large cavities ($R_{\rm cav}/a\sim3$).


In calculating the dust mass we assumed that the average disk temperature was the same between circumbinary disks and disks around single stars. In our simple model, if there is no dust within $R_{\rm cav}$ of the disk center the average dust temperature will be lower, due to the loss of the highest temperature dust. In that case, our assumed dust temperature is an over-estimate of the actual disk temperature, which leads to an under-estimate of the disk mass. The right panel of Figure~\ref{figure:mdisk_fraction} shows the combined effect of the drop in temperature and the loss of disk mass on the disk flux, assuming a power-law temperature profile and completely removing all dust with $R_{\rm cav}$ of the disk center. Including the change in average disk temperature results in a more substantial decrement in disk emission, but still requires special conditions to match our observations. For large cavities (R$_{\rm cav}/a\sim$3), large binary separations ($a\sim$5--10 au), steep temperature profiles, and compact disks the decrease in flux is consistent with that observed in our sample. While we cannot rule out these conditions, the large binary separations are unlikely; the spectroscopic surveys used when constructing our sample are most sensitive to small separations \citep[$<$1 au][]{Kounkel2019}, where the decrease in disk flux from clearing is $<$40\%.


Our simple calculations indicate that, while qualitatively we expect dust mass to decrease when the binary clears out a central cavity, in detail clearing from the binary alone may not be sufficient to explain the size of the observed mass depletion. While detailed calculations, including heating associated with the binary \citep[e.g.,][]{pierens2024} and a more detailed treatment of the density profile near the binary \citep[e.g.,][]{miranda2015,hirsh2020,penzlin2024} are needed to fully quantify the discrepancy, additional sources of gravitational clearing beyond the central binary may also contribute to a depleted disk mass.  Some high-resolution observations of circumbinary disks \citep{boehler2017,kurtovic2022,martinez-brunner2022,ginski2024} show cavities that are much larger than expected from the binary alone. While a full characterization of the binary orbit is needed to accurately relate cavity size to the expected gravitational clearing \citep{ragusa2025}, the larger than expected cavities may also be a sign of an unseen massive planet. A number of circumbinary planets are at the outer edge of the region that would be gravitationally cleared by the binary \citep[e.g.,][]{kostov2020}. If these planets are sufficiently massive \citep{crida2006,thun2018,penzlin2021} then the depleted region of the disk would be larger than predicted from binary clearing alone, and if such planets were more common around binary stars than single stars, they could explain the relative depletion of circumbinary disks.  Alternatively, an unseen tertiary companion can truncate the outer edge of the disk, decreasing the overall disk flux. Greater than 90\% of spectroscopic binaries with periods less than 5 days have a tertiary companion, while this fraction drops to 30 -- 40\% for wider spectroscopic binaries \citep{tokovinin2006,laos2020}. High spatial resolution observations are needed to search for additional stellar or planetary companions. 

\subsubsection{The Role of External Radiation\label{sec:photoevaporation}}
Environment can be an important factor in driving disk evolution. In particular, high-energy radiation from massive stars can drive external photoevaporation, rapidly depleting the disk of material \citep[e.g.,][]{winter2022}. The ONC is a well-known region subject to such external photoevaporation, most dramatically seen in the tadpole-shaped shells around propylds \citep[e.g.,][]{odell1993}. In the neighborhood of $\theta^1$ Ori C, the most massive star in the Trapezium and strongest source of high-energy radiation, the radiation field ranges from 10$^6$ G$_0$\footnote{The FUV radiation field is written as a multiple of the Habing unit, G$_0$ \citep{habing1968}, which is the integrated flux between 912 and 2400 \AA, normalized to the average flux in the solar neighborhood ($1.6\times10^{-3}$ erg s$^{-1}$ cm$^{-2}$).} at 0.01 pc down to 10$^3$ G$_0$ at 1 pc \citep{anania2025}. In radiation fields of 10 G$_0$, typical of the Taurus star forming region, \citet{haworth2018} estimate mass loss rates of 10$^{-8}$ -- 10$^{-6}$ M$_{\odot}$ yr$^{-1}$, with the strongest rates for large disks around low mass stars.

In their survey of ONC sources within 0.1 pc of $\theta^1$ Ori C, \citet{eisner2018} find that disk masses are significantly depleted relative to more isolated disks. Our tight binary sample shows a similar level of disk mass depletion (Figure~\ref{figure:mass_dist}), which may be a coincidence, or may be a sign that our sample is subject to the depletion effects of external photoevaporation. Rates of external photoevaporation increase with proximity to a source of high energy radiation, but the exact distance over which photoevaporation produces significant disk depletion is not clear. \citet{mann2014} find that average disk mass steadily increases between 0.1 pc and 1 pc away from $\theta^1$ Ori C, although the large spread in disk masses makes it difficult to tell exactly how far this influence extends. \citet{anania2025}, examining the radiation field and disk masses across multiple regions, and controlling for the correlation between disk mass and stellar mass, find that disks masses significantly decrease in radiation fields above 10$^3$ G$_0$, corresponding to $<$1 pc away from $\theta^1$ Ori C. This suggests that if any significant effect from photoevaporation exists it is most likely to be found for sources within 1 pc of $\theta^1$ Ori C.

\begin{figure}
    \center
    \includegraphics[scale=1.1]{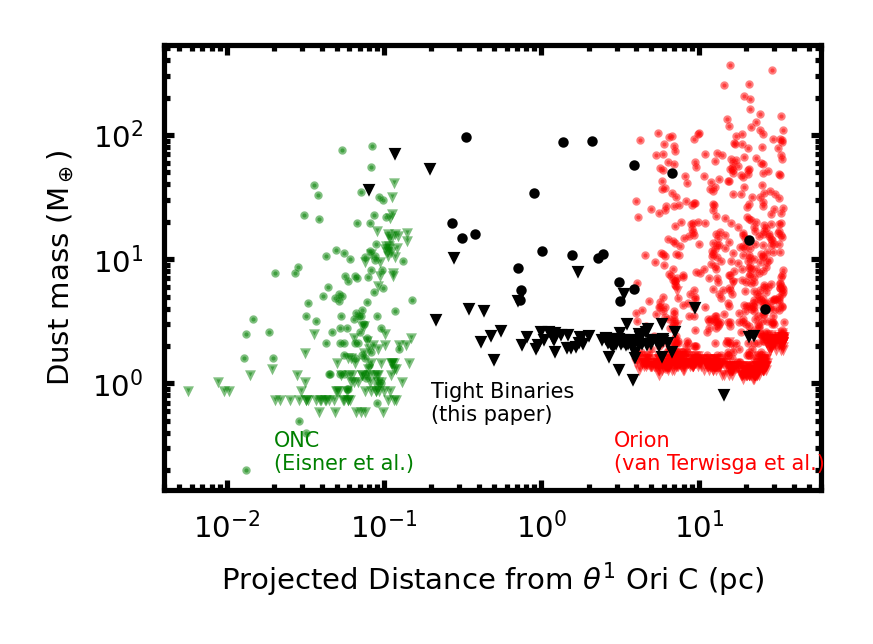}
    \includegraphics[scale=1.1]{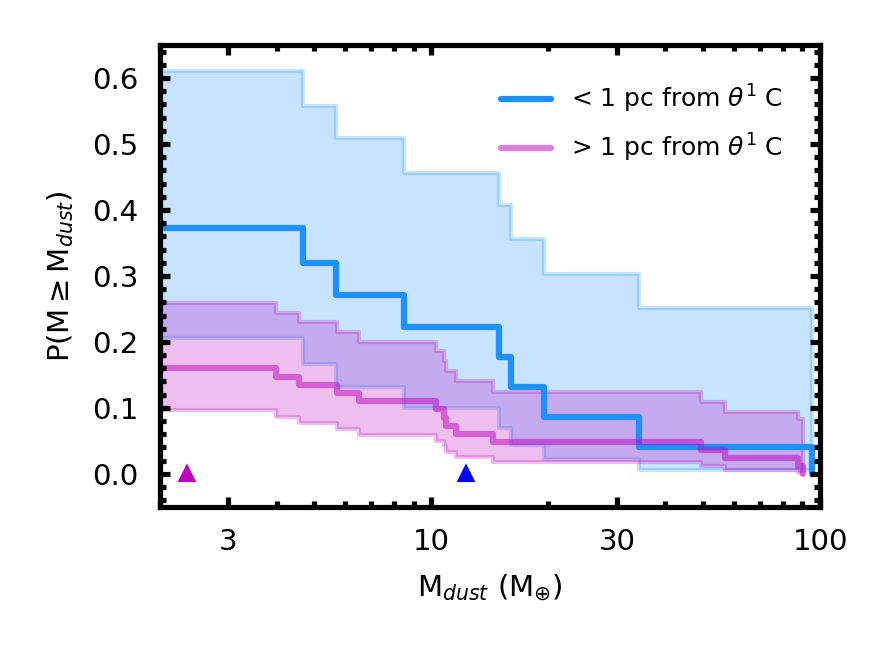}
    \caption{Top: Dust mass as a function of project distance from $\theta^1$ Ori C for the tight binaries studied here (black points), ONC targets \citep[green points,][]{eisner2018}, and targets from across the Orion cloud \citep[red points,][]{vanTerwisga2022}. The vast majority of the tight binaries fall beyond the region within 0.1 pc that \citet{eisner2018} identified as being affected by external photoevaporation. Bottom: Disk dust mass distributions comparing tight binaries within 1 pc of $\theta^1$ Ori C and more distant sources. Sources close to $\theta^1$ Ori C showing slightly higher masses, most likely due to the diminished sensitivity in a region with higher background emission; triangles indicate average upper limits for the corresponding samples.  \label{mass_theta1dist} }
\end{figure}

Figure~\ref{mass_theta1dist} shows dust mass as a function of projected distance from $\theta^1$ Ori C, comparing our tight binary sample to the ONC sample \citep{eisner2018} and the Orion sample \citep{vanTerwisga2022}. The vast majority of the tight binaries studied here are between 0.1 and 10 pc away from $\theta^1$ Ori C, at intermediate distances between that of the ONC and Orion samples. There are 22 sources in our sample, including 8 detections, between 0.1 and 1 pc, where we would expect the effect of external radiation to be the strongest. Figure~\ref{mass_theta1dist} shows the dust mass distribution for the tight binaries, splitting the sample between sources within 1 pc of $\theta^1$ Ori C and sources beyond 1 pc. There is no strong evidence that disk masses are lower closer to $\theta^1$ Ori C. If anything, the typical disk masses are higher closer to the radiation source, although this might be an observational bias; close to $\theta^1$ Ori C the background emission becomes stronger, increasing the upper limits of non-detections. The lack of a significant difference between sources $<$ 1 pc and $>$ 1 pc from $\theta^1$ Ori C suggests that external radiation is not the primary culprit for the depressed dust masses among the tight binaries, although its influence on our sample cannot be ruled out. 





\subsubsection{The Formation of Tight Binaries}

Above, we considered the effect of a binary on a disk, assuming the binary was fully in place at its present configuration, with a disk around it. Another contributing factor to the diminished disk masses may be that the tight binary formation process results in a depleted disk, leaving an imprint on our observed systems. Tight binaries likely form from the fragmentation of a massive disk rather than the fragmentation of a pre-stellar core, but in-situ disk fragmentation at $<$10 au is challenging as the high disk temperatures and large Keplerian shear at these small radii counteract gravitational fragmentation \citep{boss1986,bate1998}. Instead, a fragment may form further out in the disk and migrate inward to a final resting place close to the central star \citep{bate2002}. The migration process, and accretion onto the secondary star during migration, may sweep up a substantial amount of disk mass. \citet{offner2023}, using data from \citet{tobin2020}, show that tight binaries in class 0/I sources have disk masses less than that of single stars, consistent with a depletion happening early in the life of the binary. This fragmentation and migration mechanism can explain the excess of equal-mass binaries among tight binaries \citep{tokovinin2020}, and may contribute to the depleted disk masses that we observe.


\subsubsection{Opacity differences}
When calculating dust masses using equation~\ref{eqn:mass} we assumed that the opacity was the same between single stars and tight binaries, and that the assumption of optically thin emission was equally valid (or equally invalid) between single stars and tight binaries. Opacities depend on the complex combination of dust composition, size (or size distribution), porosity, etc. \citep[e.g.,][]{birnstiel2018}. If the grain sizes in circumbinary disks were much larger than around single stars, or the grains were more porous, or the grain size distribution was steeper, the dust opacity would be smaller \citep{birnstiel2018} than for single stars, which could account for the observed difference in the disk flux distribution. 

Narrow rings are ubiquitous in high-resolution observations of disks around single stars \citep{bae2023}, and have also been observed in a number of circumbinary disks \citep{kurtovic2022,martinez-brunner2022,long2021}. While the mechanisms behind such rings are not fully understood, a difference in the number of optically thick, narrow rings could lead to a difference in the observed disk flux. \citet{Zagaria2023} find that dust trapping at the inner edge of the circumbinary disk can lead to dust-to-gas mass ratios orders of magnitude higher than around single stars, for low-viscosity systems. If optically thick dust is a substantial component of the ALMA emission, the orientation of the disk may also play a role, since for optically thick emission the flux depends directly on the orientation of the disk. 

\subsubsection{Inclination Bias}
Inclination may also bias our sample with respect to the presence of disks. Using spectroscopy to find binaries requires moderately bright targets, half of which were observed in the optical, for sufficient SNR in the spectra. For observations with APOGEE, which make up half of our sample and are characterized in more detail than the optical sample, high SNR was achievable for targets with 7.5 $<$ H $<$ 12.5, with low SNR observations reaching down to T$_{\rm eff}\sim$3000 K \citep{Cottaar2014}. High-inclination systems with disks, or particularly young systems still embedded in an envelope, would not make it into the sample as the central sources would be too heavily extinguished.  Conversely, the spectroscopic detection of binarity is biased towards bright high-inclination systems \citep{Kounkel2019}. In essence, diskless binaries can be detected at all inclinations, including the high inclination systems where it is easiest to find binaries, but binaries with disks with high inclinations would be missed. Samples of single stars also miss systems with edge-on disks, and the exact size of the bias against edge-on circumbinary disks is difficult to characterize since the inclination at which a disk would substantially extinguish the central stars depends on the settling of dust at large radii \citep{dallesio2006}. Spatially resolved observations are needed to better understand the orientation of the observed circumbinary disks. 


\subsection{Inner-Outer disk connection}
Despite the depletion of cold dust, we find comparable rates of infrared excesses between tight binaries and single stars. Previous studies have found a depletion of infrared dust emission around tight ($<$40 au) binaries \citep{cieza2009,kraus2012}. Correlated depletion of the inner and outer disk is common among single stars \citep{hardy2015,lovell2021,pericaud2017,andrews2005}, and is indicative of rapid depletion of both the inner and outer disks.

The observed IR excesses among the binaries may reflect streamers supplied by the outer disk. The IR observations are sensitive to very small amounts of mass \citep{cieza2007}, and streamers leading from the circumbinary disk onto circumstellar disks, or directly onto the stars, are expected in binary systems \citep{artymoqicz1996,munoz2020,marzari2025} and are observed in high resolution observations in a number of systems \citep{keppler2020,alves2019}. The time-dependent infrared excess of DQ Tau, which coincides with accretion pulses from material falling onto the central stars, is consistent with material being dragged in from the outer disk during each apastron passage of this eccentric binary system \citep{mathieu1997,tofflemire2017,muzerolle2019}. In this scenario, the tentative evidence of a weaker 3-5$\mu$m excess among the sources without ALMA detections may reflect an inability of the outer disk to resupply the inner disk with material, due to the depletion or complete absence of an outer disk. 

\section{Conclusions}
We surveyed 130 spectroscopic binaries with ALMA to characterize the cold gas reservoir. Among the 107 likely cluster members, we detect significant emission from 21, with dust masses ranging from 1--100 M$_{\oplus}$. We find that the distribution of disk dust masses is lower than that around single stars within the Orion star-forming region, as well as other star-forming regions of similar ages, indicating a factor of $\sim$2 depletion in dust mass associated with the presence of a tight binary. We find that the infrared excesses are common among the binary systems, but there is evidence that the 3-5$\mu$m excesses are weaker among binary systems without ALMA detections. 

These observations indicate that the planet forming environment around tight ($<$10 au) binaries is disrupted relative to that around single stars. Taken at face value, this implies less material available to form planets, which in turn may lead to a less massive population of planets around tight binaries. Preliminary evidence suggests that circumbinary planets tend to be Saturn-sized or smaller \citep{orosz2012b,schwamb2013}, which may be related to the presence of smaller disks, but more work is needed to fully characterize this population. 

The depletion seen in the ALMA observations is difficult to explain by clearing alone. It would require systems with separations close to 10 au, compact initial disks, and steep temperature profiles. This specific set of circumstances is unlikely, especially given that spectroscopic surveys are biased towards very tight ($<$1 au) binaries. Instead, the means of uniting the stars in a tight binary, through gravitational fragmentation of a massive disk and inward migration of the secondary, may deplete the disk. Other factors, such as the presence of a tertiary stellar/planetary companion, external photoevaporation, a bias away from edge-on disks, or systematic differences in the temperature, opacity, and/or optical depth in circumbinary disks as compared to disks around single stars may also contribute to the observed depletion.

\vspace{5mm}
\facilities{ALMA}

\software{astropy \citep{2018AJ....156..123A}, casa \citep{casa}, lifelines \citep{davidson_pilon_cameron2022}, isochrones \citep{isochrones}, Multinest \citep{buchner2014}}

\begin{acknowledgments}
We thank the referee for their detailed comments that have improved the presentation and analysis in the paper. A.M.H. acknowledges support from the National Science Foundation under Grant No. AST-2307920. This paper makes use of the following ALMA data: ADS/JAO.ALMA\#2017.1.01639.S. ALMA is a partnership of ESO (representing its member states), NSF (USA) and NINS (Japan), together with NRC (Canada), MOST and ASIAA (Taiwan), and KASI (Republic of Korea), in cooperation with the Republic of Chile. The Joint ALMA Observatory is operated by ESO, AUI/NRAO and NAOJ. The National Radio Astronomy Observatory is a facility of the National Science Foundation operated under cooperative agreement by Associated Universities, Inc. This publication makes use of data products from the Wide-field Infrared Survey Explorer, which is a joint project of the University of California, Los Angeles, and the Jet Propulsion Laboratory/California Institute of Technology, funded by the National Aeronautics and Space Administration. This publication makes use of data products from the Two Micron All Sky Survey, which is a joint project of the University of Massachusetts and the Infrared Processing and Analysis Center/California Institute of Technology, funded by the National Aeronautics and Space Administration and the National Science Foundation. This work has made use of data from the European Space Agency (ESA) mission {\it Gaia} (\url{https://www.cosmos.esa.int/gaia}), processed by the {\it Gaia} Data Processing and Analysis Consortium (DPAC,
\url{https://www.cosmos.esa.int/web/gaia/dpac/consortium}). Funding for the DPAC has been provided by national institutions, in particular the institutions
participating in the {\it Gaia} Multilateral Agreement.
\end{acknowledgments}

\appendix
\section{Data Tables and Figures\label{sec:full_data}}
\begin{figure*}
    \centering
    \includegraphics[scale=2.2]{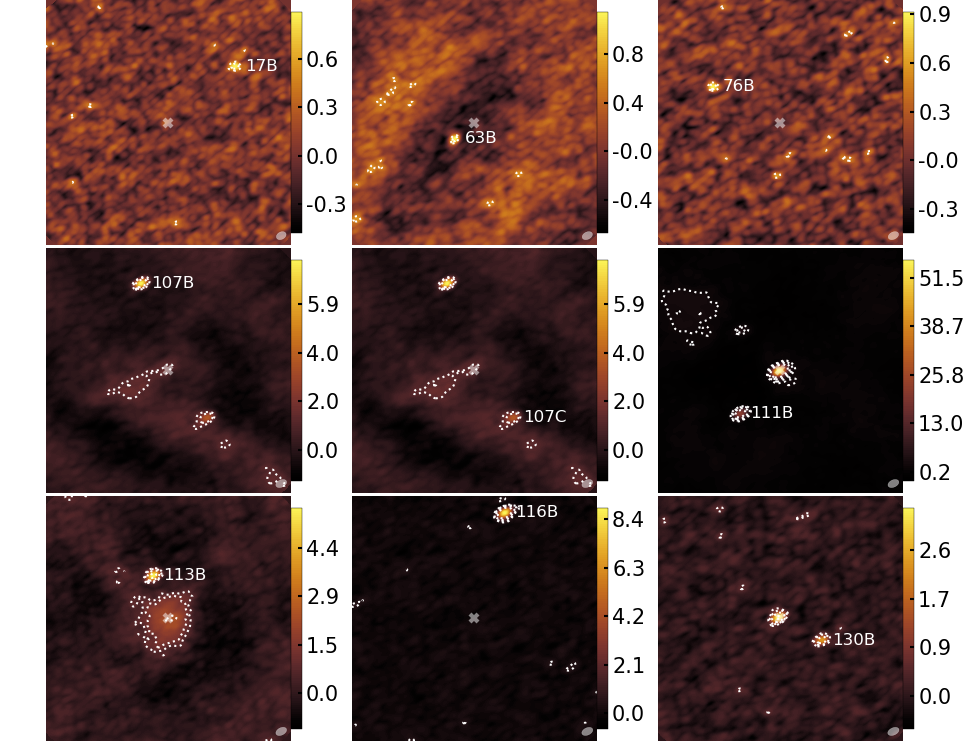}
    \caption{Continuum images, centered on the detections, of non-cluster members. Each panel is 25\arcsec on a side, with the scale bar in units of mJy bm$^{-1}$. Dotted contours are at 3, 5, 10, and 20$\sigma$ and the ellipse in the lower right is the beam size.}
    \label{fig:cont_images_excluded}
\end{figure*}

Here we show:
\begin{itemize}
    \item A table containing stellar information and measurements from the ALMA observations for cluster members (Table~\ref{table:continuum_data}).
    \item A table including ALMA observations of HCO$^{+}$ among cluster members with significant ALMA continuum emission (Table~\ref{table:hco}).
    \item Continuum images of non-cluster members with significant ALMA emission (Figure~\ref{fig:cont_images_excluded}).
    \item A table containing stellar information and measurements from the ALMA observations for non-cluster members (Table~\ref{table:continuum_data_excluded}).
    \item A table including ALMA observations of HCO${+}$ among non-cluster members with significant ALMA continuum emission (Table~\ref{table:hco-excluded}).
    
\end{itemize}

\clearpage

\startlongtable
\begin{deluxetable*}{cccccccccc}
\tablecaption{Continuum Observations\label{table:continuum_data}}
\tablewidth{0pt}
\tabletypesize{\scriptsize}
\tablehead{
\colhead{ID} & \colhead{RA} & 
\colhead{Dec} & \colhead {Other name} & \colhead{Flux} & 
\colhead{M$_{\rm dust}$}  & 
\colhead{T$_{\rm eff}$} & \colhead{$\log$~g} & 
\colhead{$p$} & \colhead{Reference}  \\ 
\colhead{} & \colhead{(hh:mm:ss)} & \colhead{(dd:mm:ss)} & & \colhead{(mJy)} & 
\colhead{(M$_{\oplus}$)} & \colhead{(K)} &
\colhead{(cm s$^{-2}$)} & \colhead{(mas)} & \colhead{} 
} 
\startdata
1 & 05:32:16.83 & -05:48:23.7 & BD-05 1284 & $<$0.59 & $<$1.66 & 7144 & 3.90 & 2.75$\pm$0.08 & F17 \\ 
2 & 5:34:23.86 & -05:15:40.3 & V1676 Ori &  $<$0.655 & $<$2.15 & 3854 & 3.62 & 2.557$\pm$0.029 & F17 \\ 
4 & 05:34:55.63 & -06:01:03.6 & V1710 Ori & $<$0.64 & $<$2.11 & 3820 & 4.49 & 2.553$\pm$0.028 & F17 \\ 
5 & 05:35:03.26 & -04:49:20.9 & V1718 Ori & 1.74$\pm$0.073 & 5.73$\pm$0.24 & 4330 & 4.34 & 2.55\tablenotemark{a} & F17 \\ 
6 & 05:35:03.92 & -05:29:03.3 & V1481 Ori & 1.73$\pm$0.069 & 5.75$\pm$0.26 & 3732 & 3.60 & 2.537$\pm$0.026 & F17 \\ 
7 & 05:34:52.49 & -04:49:40.4 & 2MASS J05345249-0449404 &$<$0.74 & $<$2.48 & 3060 & 4.71 & 2.533$\pm$0.063 & F17 \\ 
9 & 05:35:15.8 & -06:46:29.1 & 2MASS J05351580-0646291  & $<$0.62 & $<$4.10 & 5640 & 5.12 & 1.8$\pm$0.088 & F17 \\ 
11 & 05:34:52.49 & -04:49:40.4 & 2MASS J05345249-0449404 & $<$0.64 & $<$1.62 & 3770 & 4.11 & 2.915$\pm$0.182 & F17 \\ 
12 & 05:36:47.17 & -05:22:50.0 & V2706 Ori & $<$0.71 & $<$2.32 & 3260 & 3.74 & 2.561$\pm$0.038 & F17 \\ 
14 & 05:41:24.51 & -08:56:32.0 & Kiso A-1048 43 & 0.86$\pm$0.07 & 3.99$\pm$0.33 & 4490 & 4.33 & 2.15$\pm$0.02 & F17 \\ 
15 & 05:34:06.93 & -04:39:30.3 & Brun 181 & $<$0.68 & $<$2.27 & 4395 & 4.05 & 2.534$\pm$0.017 & F17 \\ 
17 & 05:34:52.2 & -04:40:11.7 & SX Ori & $<$0.59 & $<$2.12 & 5290 & 5.11 & 2.445$\pm$0.021 & F17 \\ 
18 & 05:35:11.13 & -05:36:51.1 & V2212 Ori & $<$0.62 & $<$1.98 & 4050 & 3.93 & 2.591$\pm$0.029 & F17 \\ 
19 & 05:35:15.54 & -05:25:14.0 & V1501 Ori & $<$0.97 & $<$3.32 & 4230 & 3.99 & 2.505$\pm$0.051 & F17 \\ 
20 & 05:35:28.13 & -05:23:06.4 & V2471 Ori & 25.66$\pm$0.07 & 96.4$\pm$13.0 & 3610 & 3.7 & 2.41$\pm$0.16 & F17 \\ 
21 & 05:35:31.54 & -05:40:27.8 & V1560 Ori & $<$0.69 & $<$2.49 & 4460 & 4.2 & 2.439$\pm$0.089 & F17 \\ 
22 & 05:35:36.45 & -04:21:30.4 & 2MASS J05353645-0421304 & $<$0.61 & $<$2.64 & 3820 & 4.44 & 2.226$\pm$0.214 & F17 \\ 
23 & 05:35:54.08 & -05:28:32.7 & 2MASS J05355408-0528327 & $<$0.76 & $<$1.82 & 4230 & 4.61 & 2.997$\pm$1.04 & F17 \\ 
24 & 05:36:01.85 & -05:17:36.5 & V578 Ori & $<$0.65 & $<$1.98 & 4410 & 4.32 & 2.656$\pm$0.014 & F17 \\ 
27 & 05:34:27.67 & -05:37:19.2 & V1962 Ori & 25.16$\pm$0.07 & 89.8$\pm$11.8 & 3690 & 3.52 & 2.47$\pm$0.161 & F17 \\ 
28 & 05:35:01.38 & -06:15:17.5 & Parenago 1729 & $<$0.71 & $<$2.34 & 4130 & 3.86 & 2.551$\pm$0.015 & F17 \\
29 & 05:35:11.21 & -05:17:20.90 & V2214 Ori & 2.43$\pm$0.08 & 8.50$\pm$0.55 & 4100 & 4.33 & 2.48$\pm$0.07 & F17 \\
30 & 05:35:12.14 & -05:31:38.8 & 2MASS J05351214-0531388 & $<$0.61 & $<$2.11 & 3320 & 3.85 & 2.49$\pm$0.09 & F17 \\
31 & 05:35:12.36 & -05:43:18.4 & V486 Ori & 3.37$\pm$0.07 & 10.31$\pm$0.25 & 4340 & 4.29 & 2.651$\pm$0.015 & F17 \\
32 & 05:35:15.61 & -05:24:03.0 & Brun 590 & $<$1.1 & $<$37.0 & 4490 & 5.17 & 2.504$\pm$0.035 & F17\\
33 & 05:35:17.55 & -04:53:03.7 & 2MASS J05351755-0453037 & $<$0.70 & $<$2.11 & 3670 & 3.93 & 2.67$\pm$0.13 & F17 \\
34 & 05:35:30.02 & -04:34:27.6 & 2MASS J05353002-0434276 & $<$0.64 & $<$2.05 & 3000 & 4.43 & 2.59$\pm$0.06 & F17 \\
35 & 05:35:43.11 & -04:24:55.6 & BO Ori & 1.47$\pm$0.07 & 4.93$\pm$0.25 & 4240 & 4.43 & 2.53$\pm$0.02 & F17 \\
36 & 05:35:44.63 & -04:50:09.8 & V1580 Ori & $<$0.64 & $<$2.10 & 3950 & 4.58 & 2.557$\pm$0.027 & F17 \\
38 & 05:35:17.98 & -06:04:43.0 & Parenago 1943 & $<$0.64 & $<$2.06 & 4610 & 4.69 & 2.572$\pm$0.016 & F17 \\
40 & 05:35:43.37 & -06:22:19.5 & V810 Ori & $<$0.58 & $<$1.81 & 3220 & 3.19 & 2.623$\pm$0.053 & F17 \\
41 & 05:36:17.23 & -06:17:24.5 & V2657 Ori & $<$0.66 & $<$2.17 & 3070 & 4.41 & 2.55\tablenotemark{a} & F17 \\
44 & 05:38:01.6 & -07:22:21.8 & 2MASS J05380160-0722218 & $<$0.75 & $<$0.83  & 5080 & 5.0 & 4.417$\pm$0.065 & F17 \\ 
49 & 05:41:33.38 & -07:59:56.2 & 2MASS J05413338-0759562 & 6.53$\pm$0.08 & 14.38$\pm$2.50 & 3863 & 3.29 & 3.16$\pm$0.26 & F17 \\
50 & 05:42:01.51 & -07:56:15.3 & 2MASS J05420151-0756153 & $<$0.63 & $<$2.41 & 3510 & 4.21 & 2.37$\pm$0.06 & F17 \\
51 & 05:42:23.93 & -08:09:45.9 & 2MASS J05422393-0809459 & $<$0.67 & $<$2.49 & 4480 & 4.55 & 2.403$\pm$0.029 & F17 \\
53 & 05:34:45.56 & -05:29:20.9 & V1445 Ori & $<$0.76 & $<$2.64 & 3610 & 3.81 & 2.488$\pm$0.064 & F17 \\
54 & 05:34:54.31 & -04:54:12.9 & 2MASS J05345431-0454129 & $<$0.69 & $<$2.27 & 4020 & 4.12 & 2.555\tablenotemark{a} & F17 \\
55 & 05:35:13.52 & -05:27:28.6 & JW 415 & $<$0.79 & $<$2.46 & 3980 & 4.28 & 2.626$\pm$0.073 & F17 \\
56 & 05:35:21.47 & -05:57:42.1 & Haro 4-359 & $<$0.59 & $<$1.80 & 3880 & 4.65 & 2.653$\pm$0.089 & F17 \\ 
57 & 05:35:42.61 & -05:26:08.3 & JW 964 & $<$0.67 & $<$2.44 & 3090 & 3.53 & 2.430$\pm$0.079 & F17 \\
59 & 05:34:39.03 & -04:55:28.8 & 2MASS J05343903-0455288 & $<$0.68 & $<$2.28 & 3732 & 3.35 & 2.530$\pm$0.036 & F17 \\
60 & 05:34:48.23 & -04:47:40.1 & V1703 Ori & $<$0.61 & $<$2.06 & 4010 & 4.23 & 2.52$\pm$0.02 & F17 \\ 
61 & 05:35:02.0 & -05:20:55.0 & V2118 Ori & $<$0.66 & $<$1.57 & 4200 & 4.37 & 3.00$\pm$0.67 & F17 \\
62 & 05:35:04.63 & -05:09:55.7 & Haro 4-342 & 3.64$\pm$0.07 & 10.77$\pm$0.35 & 3660 & 4.03 & 2.694$\pm$0.036 & F17 \\ 
63 & 05:35:23.49 & -05:20:01.60 & 2MASS J05352349-0520016 & $<$1.2 & $<$3.88 & 4010 & 4.13 & 2.55\tablenotemark{a} & F17 \\
64 & 05:36:37.08 & -05:04:41.46 & V657 Ori & 1.92$\pm$0.07 & 6.52$\pm$0.51 & 3681 & 3.9 & 2.521$\pm$0.083 & F17 \\
65 & 05:36:50.78 & -04:59:33.4 & HD 294267 & $<$0.64 & $<$1.09 & 5860 & 5.1 & 3.556$\pm$0.028 & F17 \\
67 & 05:34:30.207 & -04:58:30.42 & V1434 Ori & $<$0.60 & $<$1.31 & 3499 & 4.48 & 3.14$\pm$0.15 & K16 \\ 
68 & 05:34:40.87 & -05:22:42.24 & IX Ori & 5.93$\pm$0.07 & 11.68$\pm$1.27 & 4220 & 4.74 & 3.32$\pm$0.17 & K16 \\ 
69 & 05:34:52.763 & -05:00:50.91 & 2MASS J05345276-0500509 & $<$0.69 & $<$2.16 & 3220 & 4.78 & 2.622$\pm$0.043 & K16 \\
70 & 05:34:59.01 & -05:44:29.50 & KT Ori B & 3.19$\pm$0.07 & 5.79$\pm$0.18 & 4150 & \ldots & 2.502$\pm$0.038 & K16 \\ 
71 & 05:35:12.59 & -05:23:44.16 & LV Ori & $<$22.4 & $<$71.2 & 4380 & 4.04 & 2.598$\pm$0.021 & K16 \\
72 & 05:35:16.408 & -04:58:02.0 & 2MASS J05351640-0458020 & $<$0.72 & $<$2.27 & 4000 & \ldots & 2.609$\pm$0.051 & K16 \\
73 & 05:35:17.92 & -05:15:32.76 & V1334 Ori & 8.46$\pm$0.07 & 33.86$\pm$4.89 & 4370 & 4.02 & 2.329$\pm$0.162 & K16 \\ 
74 & 05:35:23.23 & -04:43:02.90 & V1736 Ori & $<$0.82 & $<$2.65 & 3935 & 3.81 & 2.578$\pm$0.022 & K16 \\
75 & 05:35:25.676 & -04:57:18.35 & 2MASS J05352567-0457183 & $<$0.64 & $<$2.11 & 4000 & \ldots & 2.551$\pm$0.056 & K16 \\
76 & 05:35:26.589 & -04:56:06.71 & V563 Ori &$<$0.70 & $<$2.32 & 3980 & 4.29 & 2.543$\pm$0.027 & K16 \\
77 & 05:35:29.307 & -05:45:38.17 & V1551 Ori & $<$0.70 & $<$2.36 & 4000 & \ldots & 2.524$\pm$0.057 & K16 \\
78 & 05:35:32.34 & -05:18:07.80 & V1563 Ori & $<$0.65 & $<$2.09 & 4190 & 4.09 & 2.582$\pm$0.054 & K16 \\
79 & 05:36:05.655 & -05:52:13.08 & 2MASS J05360565-0552130 & $<$0.67 & $<$2.11 & 3620 & 4.78 & 2.612$\pm$0.042 & K16 \\
80 & 05:33:29.386 & -05:07:49.1 & 2MASS J05332938-0507491 & $<$0.77 & $<$3.08 & 4700 & \ldots & 2.317$\pm$0.052 & K16 \\
81 & 05:33:35.714 & -05:09:23.5 & Haro 4-294 &$<$0.68 & $<$2.22 & 3531 & 3.64 & 2.567$\pm$0.044 & K16 \\
82 & 05:33:36.441 & -04:49:49.02 & ESO-HA 835 & $<$0.87 & $<$2.81 & 3750 & \ldots & 2.580$\pm$0.047 & K16 \\
83 & 05:34:27.836 & -05:43:31.52 & 2MASS J05342783-0543315 & $<$0.65 & $<$1.34 & 3560 & 4.42 & 2.897$\pm$0.035 & K16 \\
84 & 05:34:28.52 & -05:24:57.96 & V976 Ori & 2.47$\pm$0.07 & 8.76$\pm$0.27 & 4340 & 4.47 & 2.461$\pm$0.016 & K16 \\
85 & 05:34:35.681 & -05:35:52.09 & 2MASS J05343568-0535520 & $<$0.67 & $<$2.07 & 4200 & \ldots & 2.618$\pm$0.048 & K16 \\
86 & 05:34:49.98 & -05:18:44.64 & V2056 Ori & $<$0.59 & $<$1.93 & \ldots & \ldots & 2.562$\pm$0.017 & K16 \\
88 & 05:34:52.339 & -05:30:08.01 & 2MASS J05345233-0530080 & $<$0.66 & $<$2.29 & 3150 & 4.61 & 2.489$\pm$0.068 & K16 \\
89 & 05:34:53.12 & -05:47:42.5 & ESO-HA 1054 & $<$0.72 & $<$1.99 & 5000 & \ldots & 2.784$\pm$0.086 & K16 \\
90 & 05:34:57.452 & -05:30:42.02 & V1316 Ori & $<$0.75 & $<$2.64 & 3990 & 4.24 & 2.469$\pm$0.057 & K16 \\
91 & 05:35:02.497 & -05:33:09.95 & V786 Ori & $<$0.72 & $<$2.40 & 3960 & 4.2 & 2.537$\pm$0.051 & K16 \\
92 & 05:35:02.84 & -05:51:03.1 & V1716 Ori & $<$0.68 & $<$2.12 &  3700 & \ldots & 2.627$\pm$0.033 & K16 \\
93 & 05:35:03.036 & -04:59:59.8 & 2MASS J05350303-0459597 & $<$0.67 & $<$2.20 & 3210 & 4.72 & 2.559$\pm$0.038 & K16 \\
94 & 05:35:04.76 & -05:17:42.1 & V2144 Ori & 1.41$\pm$0.07 & 4.66$\pm$0.24 & 4390 & 4.45 & 0.409$\pm$0.359 & K16 \\ 
95 & 05:35:05.63 & -05:25:18.52 & LL Ori & 4.85$\pm$0.07 & 1.63$\pm$0.23 & 4640 & 4.4 & 2.547$\pm$0.02 & K16 \\
96 & 05:35:06.08 & -05:52:39.3 & 2MASS J05350608-0552393 & $<$0.66 & $<$5.34 & 3700 & \ldots & 1.629$\pm$0.37 & K16 \\ 
97 & 05:35:06.59 & -05:59:51.4 & 2MASS J05350658-0559514 & $<$0.64 & $<$2.23 & \ldots & \ldots & 2.483$\pm$0.073 & K16 \\
98 & 05:35:06.828 & -05:10:38.53 & V1721 Ori & $<$0.71 & $<$2.51 & 3605 & 3.70 & 2.465$\pm$0.043 & K16\\
99 & 05:35:07.24 & -04:50:25.4 & 2MASS J05350724-0450254 & $<$0.66 & $<$2.09 & 3411 & 3.67 & 2.606$\pm$0.05 & K16 \\
100 & 05:35:09.96 & -05:57:11.9 & ESO-HA 1121 & 17.60$\pm$0.08 & 56.9$\pm$2.3 & 4070 & 3.96 & 2.579$\pm$0.054 & K16 \\
101 & 05:35:11.97 & -05:20:33.1 & V2221 Ori & $<$1.2 & $<$4.20 & \ldots & \ldots & 2.476$\pm$0.057 & K16 \\
102 & 05:35:12.9 & -05:59:38.5 & 2MASS J05351290-0559384 & $<$0.68 & $<$2.12 & 3960 & 4.52 & 2.624$\pm$0.029 & K16 \\
103 & 05:35:14.668 & -05:08:52.08 & 2MASS J05351466-0508520 & $<$0.63 & $<$1.98 & 3451 & 3.60 & 2.607$\pm$0.039 & K16 \\
104 & 05:35:15.758 & -05:21:39.82 & V1332 Ori & $<$15.8 & $<$54.3 & \ldots & \ldots & 2.50$\pm$0.08 & K16 \\
105 & 05:35:17.139 & -05:12:39.41 & 2MASS J05351713-0512394 & $<$0.77 & $<$2.60 & 4200 & \ldots & 2.52$\pm$0.08 & K16 \\
106 & 05:35:18.098 & -04:31:19.33 & V2322 Ori & $<$0.75 & $<$2.24 & 3850 & \ldots & 2.684$\pm$0.045 & K16 \\
107 & 05:35:20.04 & -05:21:05.9 & V491 Ori & $<$2.37 & $<$10.4 & 4410 & 4.01 & 2.21$\pm$0.16 & K16 \\ 
108 & 05:35:22.817 & -05:44:42.89 & V1529 Ori & $<$0.72 & $<$2.29 & 3499 & 3.64 & 2.597$\pm$0.038 & K16 \\ 
109 & 05:35:23.97 & -05:59:41.9 & Parenago 2024 & $<$0.68 & $<$2.56 & 3999 & 3.88 & 2.390$\pm$0.028 & K16 \\ 
110 & 05:35:24.25 & -05:25:18.48 & V2421 Ori & 4.06$\pm$0.07 & 14.94$\pm$0.67 & \ldots & \ldots & 2.418$\pm$0.049 & K16 \\
111 & 05:35:24.43 & -05:24:39.8 & V2423 Ori & 56.60$\pm$0.07 & 194.4$\pm$9.4 & 4120 & 4.15 & 2.498$\pm$0.059 & K16 \\
112 & 05:35:25.4 & -05:51:08.64 & BD-05 1321 & 1.47$\pm$0.07 & 4.58$\pm$0.23 & 6220 & 3.92 & 2.626$\pm$0.021 & K16\\
113 & 05:35:26.06 & -05:08:37.9 & 2MASS J05352606-0508379 & $<$2.61 & $<$8.1 & 3850 & \ldots & 2.632$\pm$0.081 & K16 \\
114 & 05:35:26.417 & -05:55:26.68 & 2MASS J05352641-0555266 & $<$0.74 & $<$2.43 & 3750 & \ldots & 2.528$\pm$0.066 & K16 \\
115 & 05:35:30.48 & -05:24:23.0 & Brun 750 & $<$0.65 & $<$2.21 & 4310 & 4.24 & 2.514$\pm$0.097 & K16 \\ 
116 & 05:35:30.913 & -05:18:17.95 & 2MASS J05353091-0518179 & $<$0.76 & $<$4.67 & 3690 & 4 & 1.869$\pm$0.115 & K16 \\
117 & 05:35:32.52 & -05:26:10.4 & V1565 Ori & $<$0.69 & $<$2.70 & 4090 & 4.34 & 2.342$\pm$0.079 & K16 \\ 
118 & 05:35:33.14 & -05:47:07.44 & V1744 Ori & $<$0.68 & $<$2.13 & 3720 & 4.28 & 2.616$\pm$0.049 & K16 \\
119 & 05:35:33.456 & -04:56:01.76 & V1745 Ori & $<$0.79 & $<$2.60 & 3750 & \ldots & 2.554$\pm$0.057 & K16 \\
121 & 05:35:34.093 & -04:32:37.01 & ESO-HA 1296 & $<$0.81 & $<$3.08 & \ldots & \ldots & 2.375$\pm$0.059 & K16 \\
122 & 05:35:37.341 & -06:00:00.21 & V1749 Ori & $<$0.78 & $<$2.40 & 2310 & 4.32 & 2.637$\pm$0.049 & K16 \\
123 & 05:35:38.214 & -05:14:19.02 & 2MASS J05353821-0514190 & $<$0.71 & $<$2.28 & 3760 & 3.85 & 2.588$\pm$0.051 & K16 \\
124 & 05:35:39.07 & -05:08:56.4 & V1576 Ori & $<$0.75 & $<$2.44 & 4390 & 3.99 & 2.571$\pm$0.015 & K16 \\
125 & 05:35:39.495 & -04:40:19.42 & 2MASS J05353949-0440194 & $<$0.72 & $<$2.26 & 4180 & 4.12 & 2.613$\pm$0.038 & K16 \\
127 & 05:35:43.24 & -05:09:17.1 & V569 Ori & $<$0.72 & $<$2.37 & 4200 & 4.12 & 2.55\tablenotemark{a} & K16 \\
128 & 05:35:48.853 & -05:00:28.56 & V1584 Ori & $<$0.72 & $<$2.35 & 4250 & \ldots & 2.566$\pm$0.043 & K16 \\
129 & 05:36:03.355 & -04:57:40.48 & 2MASS J05360335-0457404 & $<$0.64 & $<$2.09 & 4000 & \ldots & 2.566$\pm$0.039 & K16 \\
130 & 05:36:06.61 & -05:41:54.23 & 2MASS J05360660-0541543 & 3.31$\pm$0.08 & 11.10$\pm$0.83 & \ldots & \ldots & 2.538$\pm$0.089 & K16 \\
\enddata
\tablenotetext{a}{No {\it Gaia} detection. Parallax is assumed to be the cluster average.}
\end{deluxetable*}

\clearpage

\begin{deluxetable*}{ccc}
\tablecaption{HCO$^+$ observations - Cluster members\label{table:hco}}
\tablehead{\colhead{Object Number} & \colhead{HCO$^+$ integrated flux (Jy km s$^{-1}$)} & \colhead{Gas Mass (M$_{jup}$)}}
\startdata
5 & $<$0.18 & $<$1.4\\
6 & $<$0.18 & $<$1.4\\
14 & $<$0.17 & $<$1.9\\
20 & 0.74$\pm$0.06 & 6.4$\pm$0.9\\
27 & 0.66$\pm$0.07 & 5.5$\pm$0.9\\
29 & $<$0.21 & $<$1.7\\
31 & $<$0.17 & $<$1.3\\
35 & $<$0.21 & $<$1.6\\
49 & $<$0.21 & $<$1.0\\
62 & $<$0.18 & $<$1.2\\
63 & $<$0.44 & $<$3.4\\
64 & $<$0.19 & $<$1.5\\
68 & $<$0.19 & $<$0.9\\
70 & $<$0.20 & $<$1.6\\
73 & $<$0.21 & $<$1.9\\
84 & 0.65$\pm$0.07 & $<$5.5$\pm$0.6\\
94 & $<$0.21 & $<$1.6\\
95 & $<$0.21 & $<$1.6\\
100 & 0.27$\pm$0.05 & 2.0$\pm$0.4\\
110 & $<$0.41 & $<$3.5\\
111 & 1.87$\pm$0.05 & 15.1$\pm$0.8\\
112 & $<$0.20 & $<$1.4\\
130 & $<$0.20 & $<$1.6\\
\enddata
\end{deluxetable*}

\begin{rotatetable}
\startlongtable
\begin{deluxetable*}{ccccccccccc}
\tablecaption{Continuum observations - non-cluster members\label{table:continuum_data_excluded}}
\tablewidth{700pt}
\tabletypesize{\scriptsize}
\tablehead{
\colhead{ID} & \colhead{RA} & \colhead{Dec} & 
\colhead{Flux} & \colhead{T$_{eff}$} & 
\colhead{$\log g$} & \colhead{p} &  \colhead{$\mu_{\alpha}$, $\mu_{\delta}$} & \colhead{RV} & \colhead{Reference} & \colhead{Reason for exclusion}\\ 
\colhead{} & \colhead{(hh:mm:ss)} & \colhead{dd:mm:ss} & \colhead{(mJy)} & \colhead{(K)} &
\colhead{(cm s$^{-2}$)} & \colhead{(mas)} & \colhead{(mas yr$^{-1}$)} & \colhead{km s$^{-1}$} & \colhead{} & \colhead{}} 
\startdata
3 & 05:35:28.6 & -04:55:03 & $<$0.595  & 4700 & 4.58 & 3.205$\pm$0.014 & 7.66$\pm$0.01, -11.03$\pm$0.01 & 29.08$\pm$0.75 & F17 & proper motion \\
8 & 05:35:10.77 & -06:51:55.7 & $<$0.65 & 6510 & 5.99 & 2.476$\pm$0.016 & 69.67$\pm$0.01, -7.03$\pm$0.01 & 102.1$\pm$1.5 & F17 & proper motion \\
10 & 05:43:24.64 & -08:13:27.5 & $<$0.64 & 3830 & 5.18 & 5.7$\pm$0.026 & -5.33$\pm$0.02, 5.42$\pm$0.02 & 38.1$\pm$0.5 & F17 & parallax, proper motion\\
13 & 05:38:25.89 & -06:59:51.9 & $<$0.79 & 3820 & 5.39 & 5.676$\pm$0.037 & -12.98$\pm$0.04, -13.39$\pm$0.03 & -0.72$\pm$1.02 & F17 & parallax, proper motion\\
16 & 05:34:30.87 & -04:25:07.0 & $<$0.62 & 4300 & 5.1 & 2.565$\pm$0.033 & 21.32$\pm$0.03, -25.07$\pm$0.02 & -67.67$\pm$0.97 & F17 & proper motion\\
17B & 05:34:51.74 & -04:40:05.94 & 0.87$\pm$0.07 & \ldots & \ldots & 2.591$\pm$0.029 & \ldots & \ldots & \ldots &  off-center \\
25 & 05:32:44.07 & -05:29:52.3 & $<$0.68 & 4770 & 4.73 & 2.287$\pm$0.094 & 4.11$\pm$0.08, -13.46$\pm$0.07 & 22.63$\pm$1.05 & F17 & proper motion \\
26 & 05:33:16.24 & -06:13:19.5 & $<$0.60 & 4375 & 4.48 & 3.864$\pm$0.014 & -13.70$\pm$0.01, -0.16$\pm$0.01 & 18.33$\pm$0.98 & F17 & parallax, proper motion\\
37 & 05:34:26.26 & -06:46:09.3 & $<$0.72 & 4450 & 5.39 & 2.953$\pm$0.024 & -7.92$\pm$0.02, 10.04$\pm$0.02 & -31.2$\pm$0.5 & F17 & proper motion\\
39 & 05:35:29.92 & -06:44:15.1 & $<$0.69 & 5870 & 5.26 & 1.844$\pm$0.032 & 16.55$\pm$0.03, -11.45$\pm$0.03 & 40.6$\pm$2.6 & F17 & proper motion\\
42 & 05:37:01.0 & -07:35:46.9 & $<$0.79 & 3150 & 5.16 & 12.602$\pm$0.031 & 27.61$\pm$0.03, -13.06$\pm$0.03 & 25.1$\pm$1.6 & F17 & parallax, proper motion\\
43 & 05:37:11.61 & -07:23:23.9 & $<$0.69 & 5060 & 4.64 & 2.517$\pm$0.013 & -5.68$\pm$0.01, 13.09$\pm$0.01 & 79.4$\pm$1.8 & F17 & proper motion\\
45 & 05:40:20.47 & -09:11:57.2 & $<$0.82 & 4680 & 4.53 & 2.45$\pm$0.02 & 13.69$\pm$0.02, -0.24$\pm$0.02 & -41.25$\pm$0.96 & F17 & proper motion \\
46 & 05:40:55.3 & -08:13:16.9 & $<$0.67 & 5450 & 5.28 & 3.013$\pm$0.014 & 4.10$\pm$0.01, -9.05$\pm$0.01 & -12.3$\pm$0.3 & F17 & proper motion\\
47 & 05:41:15.24 & -07:52:11.1 & $<$0.66 & 4660 & 4.89 & 2.943$\pm$0.019 & -6.15$\pm$0.02, -3.95$\pm$0.01 & 5.6$\pm$1.6 & F17 & proper motion\\
48 & 05:41:30.37 & -09:12:26.8 & $<$0.71 & 4040 & 5.09 & 4.464$\pm$0.026 & -18.75$pm$0.02, -9.28$\pm$0.02 & 4.7$\pm$1.9 & F17 & parallax, proper motion\\
52 & 05:34:13.47 & -04:23:53.9 & $<$0.64 & 3500 & 4.35 & 4.265$\pm$0.037 & 36.71$pm$0.03, -10.32$\pm$0.03 & 46.47$\pm$1.03 & F17 & parallax, proper motion\\
58 & 05:33:51.3 & -04:51:47.7 & $<$0.72 & 6590 & 4.52 & 0.857$\pm$0.077 & 0.40$pm$0.07, 8.62$\pm$0.06 & 81.89$\pm$4.1 & F17 & parallax, proper motion\\
63B & 05:35:23.62 & -05:20:03.26 & 1.01$\pm$0.07  & 4010 & 4.13 & \ldots & \ldots & \ldots & \ldots & off-center \\
66 & 05:34:21.59 & -05:10:13.95 & $<$0.74 & 5490 & \ldots & 1.315$\pm$0.042 & -3.01$\pm$0.04, 1.10$\pm$0.03 & 33.9$\pm$0.2 & K16 & parallax\\
70B & 05:34:59.12 & -05:44:30.85 & 2.60$\pm$0.07 & 4150 & \ldots & 2.453$\pm$0.039 & 0.36$\pm$0.04, 1.11$\pm$0.03 & \dots & K16 & off-center\\
76B & 05:35:27.05 & -04:56:03.02 & 0.94$\pm$0.07 & \ldots & \ldots & \ldots & \ldots & \ldots & \ldots & off-center\\
107B & 05:35:20.22 & -05:20:57.02 & 7.92$\pm$0.07 & \ldots & \ldots & 3.982$\pm$0.765 & -1.21$\pm$0.69, -2.36$\pm$0.56 & \ldots & \ldots & off-center\\ 
107C & 05:35:19.77 & -05:21:10.68 & 3.80$\pm$0.07 & \ldots & \ldots & 2.576$\pm$0.736  & \ldots & \ldots & \ldots & off-center\\
111B & 05:35:24.70 & -05:24:44.16 & 21.17$\pm$0.07 & 4120 & \ldots & \ldots & \ldots & \ldots & \ldots & off-center\\
113B & 05:35:26.16 & -05:08:33.49 & 5.59$\pm$0.08 & \ldots & \ldots & 2.632$\pm$0.081 & \ldots & \ldots & \ldots & off-center\\
116B & 05:35:30.70 & -05:18:07.18 & 8.92$\pm$0.08 & \ldots & \ldots & 2.462$\pm$0.059 & 0.37$\pm$0.06, -1.33$\pm$0.05 & 25.4$\pm$0.3 & \ldots & off-center\\
120 & 05:35:33.85 & -05:48:21.1 & $<$0.64 & 4500 & \ldots & 1.272$\pm$0.125 & 2.67$\pm$0.11, -5.67$\pm$0.10 & \ldots & K16 & parallax\\
126 & 05:35:42.84 & -05:51:36.8 & $<$0.70 & 6500 & 4.38 & 1.647$\pm$0.067 & 0.84$\pm$0.06, -12.49$\pm$0.05 & -0.24$\pm$2.47 & K16 & proper motion\\
130B & 05:36:06.32 & -05:41:56.56 & 2.35$\pm$0.07 & \ldots & \ldots & 2.606$\pm$0.02 & 1.38$\pm$0.02, 0.72$\pm$0.02 & 27.3$\pm$0.7 & \ldots & off-center \\
%
%
%
%
%
%
%
%
%
%
%
%
%
\enddata
\tablecomments{Selection criteria: (1) off-center: The detected source is offset from the phase center by $>$1$\arcsec$, (2) parallax: For {\it Gaia} detections with RUWE$<$1.4, we exclude sources with $p<$1.5 mas or $p>$ 3.5 mas, (3) proper motion: For {\it Gaia} detections with RUWE$<$1.4, we exclude sources with $| \mu_{\alpha}|>$ 5 mas yr$^{-1}$ or $| \mu_{\delta}|>$ 5 mas yr$^{-1}$.}
\end{deluxetable*}
\end{rotatetable}

\clearpage

\begin{deluxetable*}{cc}
\tablecaption{HCO$^+$ observations - Non-cluster members\label{table:hco-excluded}}
\tablehead{\colhead{Object Number} & \colhead{HCO$^+$ integrated flux (Jy km s$^{-1}$)}}
\startdata
17B & $<$0.17 \\
70B & $<$0.20 \\
76B & $<$0.19 \\
107B & $<$0.46 \\
107C & $<$0.46 \\
111B & 0.81$\pm$0.02 \\
113B & $<$0.32 \\
116B & $<$0.20 \\
130C & $<$0.20 \\
\enddata\end{deluxetable*}

\section{HCO$^{+}$ models\label{sec:hcoplus_models}}
To put our HCO$^{+}$ observations in context we generate simple parametric models for comparison. These models are not a fit to the data, but are approximations based on best estimates.

To model the molecular line emission, we use the parametric surface density and disc temperature structure as described in \citet{flaherty20} and references therein. To briefly summarize, the surface density is assumed to follow a power law with an exponential tail:
\begin{equation}
\Sigma_{\rm gas}(r) = \frac{M_{\rm gas}(2-\gamma)}{2\pi R^2_c}\left(\frac{r}{R_c}\right)^{-\gamma}\exp\left[-\left(\frac{r}{R_c}\right)^{2-\gamma}\right],
\end{equation}
where $M_{\rm gas}$, $R_c$, and $\gamma$ are the gas mass (in M$_{\odot}$), critical radius (in au) and power law index respectively. The disc extends from $R_{\rm in}$ to 1000 au. 

We assume the disk is vertically isothermal with a radial profile that follows a power law:
\begin{equation}
    T_{\rm gas} = T_{0}\left(\frac{r}{150\ \rm au}\right)^{q}
\end{equation}
A hydrostatic equilibrium calculation is performed at each radius to derive the gas volume density at each height above the midplane, based on the temperature and surface density structure. HCO$^{+}$ is spread throughout the disc assuming a constant abundance relative to H$_2$ of 10$^{-9}$. The velocity profile is assumed to be Keplerian, with additional corrections for the pressure support of the gas and the height above the midplane. The line profile is assumed to be a Gaussian whose width is set by thermal motion only ($\Delta V = \sqrt{\left(2 k_B T_{\rm gas}(r,z)/m_{HCO}\right)}$), with no turbulence. 

We use distances from \textit{Gaia} DR3 measurements of the parallax and the gas masses derived directly from the HCO$^{+}$. To estimate the stellar mass we use the \texttt{isochrones} package \citep{isochrones} which uses Multinest \citep{buchner2014} to estimate the stellar parameters, in this case stellar mass, based on the input observations, in this case T$_{\rm eff}$ and $\log g$, accounting for the uncertainties in the observations. We use the single star model fit, rather than accounting for a binary star, which biases our estimates towards the stellar parameters of the primary star. At most we expect this to introduce a factor of 2 error in the stellar mass (i.e., if the system is an equal mass binary). 

For a given set of model parameters, and the resulting density, temperature, velocity structure, model images are generated by solving the radiative transfer equation, as described in \citet{rosenfeld13}. Images are generated at the same velocities as the data, with Hanning smoothing applied to the model images. Model visibilities are generated using \texttt{vis\_sample} \citep{loomis2018}, and cleaned model images are generated from the model visibilities using the same cleaning prescription as applied to the data.

Table~\ref{table:common_model_params} lists the common values used across all of the models. To approximate the data we varied the gas temperature of the disk, and the appropriate values are listed in Table~\ref{table:specific_model_params}.

\begin{deluxetable}{cc}
\tablecaption{Common model parameters\label{table:common_model_params}}
\tablehead{\colhead{Parameter} & \colhead{Value}}
\startdata
[HCO$^{+}$/H$_2$] & 10$^{-9}$ \\
$q$ & -0.25 \\
$\gamma$ & 1 \\
R$_c$ & 100 au \\
R$_{in}$ & 1 au \\
$i$ & 45$^{\circ}$\\
\enddata
\end{deluxetable}

\begin{deluxetable}{ccccc}
\tablecaption{Source-specific model parameters\label{table:specific_model_params}}
\tablehead{\colhead{Source}&\colhead{M$_*$ (M$_{\odot}$)} & \colhead{M$_{\rm gas}$ (M$_{\odot}$)} & \colhead{T$_{0}$ (K)} & \colhead{v$_{sys}$ (km s$^{-1}$)}}
\startdata
20 & 0.27 & 0.006 & 23 & 25.8\\
27 & 0.32 & 0.006 & 18.5 & 24.8\\
84 & 0.65 & 0.005 & 32 & 23.7\\
100 & 0.57 & 0.002 & 17 & 25.8\\
111 & 0.61 & 0.014 & 42 & 23.6\\
\enddata
\end{deluxetable}

\bibliography{paper}{}
\bibliographystyle{aasjournal}

\end{document}